\documentclass[nofootinbib, pre, onecolumn, 12pt]{revtex4}

\usepackage[newfloat, frozencache]{minted}

\usepackage{amsmath}
\usepackage{amssymb}
\usepackage{graphicx}
\usepackage{bm}
\usepackage{braket}
\usepackage{color, soul}
\usepackage[dvipsnames]{xcolor}
\usepackage{bbm}
\DeclareMathOperator{\id}{\mathbbm{1}}
\DeclareMathOperator\arctanh{arctanh}
\graphicspath{ {./Figures/} }
\usepackage{float}
\usepackage{soul}
\usepackage{capt-of}

\newcommand{\beq}{\begin{equation}}
\newcommand{\eeq}{\end{equation}}

\newcommand{\python}[1]{\mintinline{python}|#1|}
\newcommand{\bash}[1]{\mintinline{bash}|#1|}

\newenvironment{customcode}{\VerbatimEnvironment\begin{minted}[
	bgcolor=gray!6,
	fontfamily=tt,
	gobble=0,
	fontsize=\small,
	breaklines=true,
	framerule=0.4pt,
	funcnamehighlighting=true,
	tabsize=2,
	obeytabs=false,
	mathescape=false
	samepage=true, 
	showspaces=false,
	showtabs =false,
	texcl=false,
	linenos=false,
	xleftmargin=\parindent,
	numbersep=1pt,
	texcomments
]{pycon}}{\end{minted}}

\usepackage{hyperref}

\newcommand\rurl[1]{%
  \href{https://#1}{\nolinkurl{#1}}%
}

\begin{document}

\title{Applications of Near-Term Photonic Quantum Computers: Software and Algorithms}
\author{Thomas R. Bromley}
\affiliation{Xanadu, 777 Bay Street, Toronto, Canada}
\author{Juan Miguel Arrazola}
\affiliation{Xanadu, 777 Bay Street, Toronto, Canada}
\author{Soran Jahangiri}
\affiliation{Xanadu, 777 Bay Street, Toronto, Canada}
\author{Josh Izaac}
\affiliation{Xanadu, 777 Bay Street, Toronto, Canada}
\author{Nicol\'as Quesada}
\affiliation{Xanadu, 777 Bay Street, Toronto, Canada}
\author{Alain Delgado Gran}
\affiliation{Xanadu, 777 Bay Street, Toronto, Canada}
\author{Maria Schuld}
\affiliation{Xanadu, 777 Bay Street, Toronto, Canada}
\author{Jeremy Swinarton}
\affiliation{Xanadu, 777 Bay Street, Toronto, Canada}
\author{Zeid Zabaneh}
\affiliation{Xanadu, 777 Bay Street, Toronto, Canada}
\author{Nathan Killoran}
\affiliation{Xanadu, 777 Bay Street, Toronto, Canada}

\begin{abstract}
Gaussian Boson Sampling (GBS) is a near-term platform for photonic quantum computing. Recent efforts have led to the discovery of GBS algorithms with applications to graph-based problems, point processes, and molecular vibronic spectra in chemistry. The development of dedicated quantum software is a key enabler in permitting users to program devices and implement algorithms. In this work, we introduce a new applications layer for the Strawberry Fields photonic quantum computing library. The applications layer provides users with the necessary tools to design and implement algorithms using GBS with only a few lines of code. This paper serves a dual role as an introduction to the software, supported with example code, and also a review of the current state of the art in GBS algorithms.
\end{abstract}

\maketitle

\section{Introduction}
Quantum computing is an interdisciplinary field combining techniques from physics, computer science, mathematics, and engineering~\cite{nielsen2010, ladd2010quantum}. Currently, a shift has occurred where research efforts have partially migrated to industry, owing to a desire to scale and commercialize quantum technologies~\cite{mohseni2017commercialize}. As a consequence, early quantum computers are being made available to the public, for example through cloud access to quantum hardware. Many applications of quantum computing such as factoring~\cite{shor1994algorithms, gidney2019factor} and Hamiltonian simulation~\cite{lloyd1996universal, berry2015hamiltonian} require large-scale fault-tolerant quantum computers that likely won't be available in the immediate future~\cite{acin2018quantum}. Instead, existing quantum computers have limited size, connectivity, and circuit depth~\cite{preskill2018quantum}. They may also be sub-universal, tailored to perform specific tasks~\cite{kadowaki1998quantum,johnson2011quantum,aaronson2013, wang2017high, georgescu2014quantum, bernien2017probing}.  

The challenge of identifying applications for near-term quantum computers can be partially addressed by creating tools that make it easier to develop quantum algorithms. A key enabler in this regard is the development of quantum software. It provides users with a high-level toolset to design and execute quantum algorithms without the need of expert knowledge~\cite{strawberryfields, pennylane, mcclean2017openfermion, forest, cirq, qiskit}, as well as giving access to hardware and high-performance simulators~\cite{walrus, smelyanskiy2016qhipster, gheorghiu2018quantum, zhang2019alibaba, qrack} for benchmarking and development.
Indeed, it is now understood that the existence of quality quantum software is an important ingredient for progress in quantum computing~\cite{zeng2017first, fingerhuth2018open}. 

Currently-available quantum computers are also characterized by the diversity of available physical platforms: superconducting qubits~\cite{devoret2013superconducting}, trapped ions~\cite{haffner2008quantum}, photonics~\cite{o2009photonic}, quantum annealers~\cite{das2008colloquium}, semiconductor qubits~\cite{kloeffel2013prospects}, and Rydberg atoms~\cite{adams2019rydberg} are all being pursued as paradigms for building quantum computers. Several of these approaches have been studied for decades, while others have only been discovered more recently.

Gaussian Boson Sampling (GBS) was introduced recently as a special-purpose photonic platform to perform sampling tasks that are intractable for classical computers~\cite{hamilton2017, kruse2019detailed}. It was quickly realized that GBS offers significant versatility in the scope of problems that can be encoded into the device. Indeed, several GBS algorithms have been developed, with applications to graph optimization~\cite{bradler2018gaussian, arrazola2018quantum, arrazola2018using}, molecular docking~\cite{banchi2019molecular}, graph similarity~\cite{bradler2018graph,schuld2019quantum,bradler2019duality}, point processes~\cite{jahangiri2019point}, and quantum chemistry~\cite{huh2017vibronic}. Progress has also been made in demonstrating the viability of the required sources of squeezed light~\cite{harder2013optimized, harder2016single, vernon2018scalable, vaidya2019broadband} and in reporting the first experimental implementations of GBS~\cite{sparrow2018simulating, zhong2019experimental}.

\begin{center}
\begin{figure}[t!]
\includegraphics[width= 0.75\columnwidth]{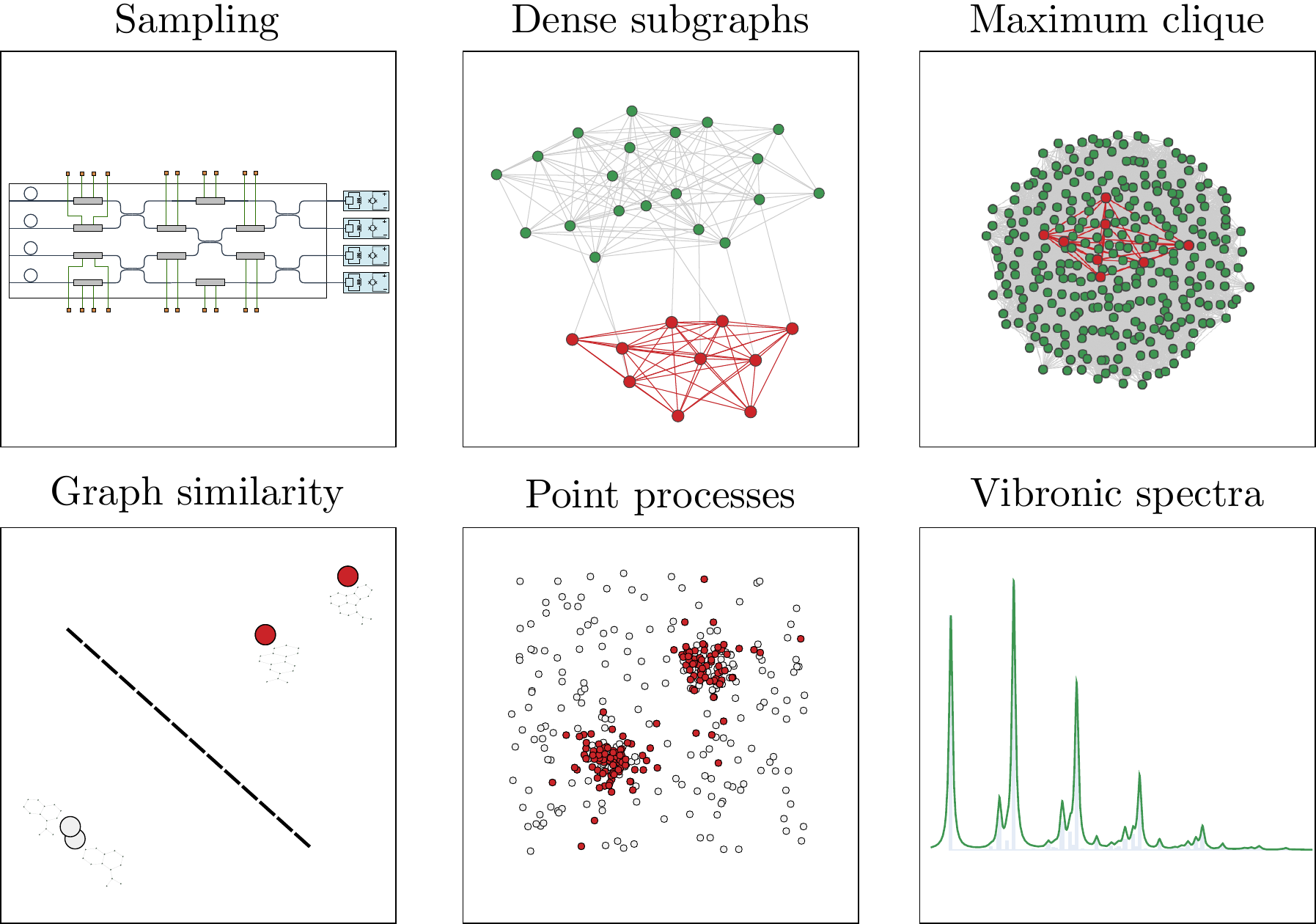}
\caption{An overview of GBS algorithms contained in the Strawberry Fields applications layer.  
}\label{Fig:Main}
\end{figure}
\end{center}

In this work, we introduce the GBS applications layer of Strawberry Fields\footnote{This document refers to Strawberry Fields version 0.12. Full documentation is available online at \rurl{strawberryfields.readthedocs.io} and the code is available at \rurl{github.com/XanaduAI/strawberryfields}.}, an open-source Python library for photonic quantum computing~\cite{strawberryfields}. The applications layer is built with the goal of providing users with the capability to implement and test GBS algorithms using only a few lines of code. Specifically, it contains modules dedicated to dense subgraph identification, maximum clique, graph similarity, point processes, and vibronic spectra. Programming of GBS devices, generating samples, and classical post-processing of the outputs are taken care of automatically by built-in functions. The applications layer also provides methods for problem visualization, a collection of pre-generated GBS datasets, and tutorials for first-time users covering each of the algorithms. An overview of GBS algorithms covered in the applications layer is shown in Fig.~\ref{Fig:Main}.

\vspace{-1cm}

In the following, we begin with a short review of GBS, with a focus on describing how to program GBS devices. We then give an overview of the applications layer, its structure, scope, and high-level functionality. The rest of the paper dedicates a section to each individual application, covering its mathematical formulation and describing the corresponding GBS algorithm. Each of these sections uses example code to outline how the algorithms are implemented in Strawberry Fields.

\section{Gaussian Boson Sampling}

In quantum optics, the fundamental physical systems of interest are optical modes of the quantized electromagnetic field, often referred to as qumodes. As opposed to qubits that are described by two-dimensional Hilbert spaces, qumodes are mathematically represented by a Hilbert space of infinite dimension. A general qubit state can be written as $\ket{\psi}=c_0\ket{0}+c_1\ket{1}$, with $|c_0|^2+|c_1|^2=1$. A general qumode state can be expressed as $\ket{\psi}=\sum_{n=0}^\infty c_n \ket{n}$, with $\sum_{n=0}^{\infty}|c_n|^2=1$. The basis states $\ket{0}, \ket{1}, \ket{2}, \ldots$ are known as \emph{Fock states}, and the Fock state $\ket{n}$ has the physical interpretation of a qumode with $n$ photons.

The state of a system of $m$ qumodes can also be uniquely specified by its Wigner function $W(\bm{q},\bm{p})$~\cite{gqi2012,FR-serafini2017}, where $\bm{q} \in \mathbb{R}^m$ and $\bm{p} \in \mathbb{R}^m$ are are known respectively as the position and momentum quadrature vectors. Gaussian states are characterized by having a Wigner function which is a Gaussian distribution. They can be elegantly described in terms of a $2m\times 2m$ covariance matrix $\bm{V}$ and two $m$-dimensional vectors of means $\bm{\bar{q}},\bm{\bar{p}}$. Alternatively, it is often more suitable to write the covariance matrix in terms of the complex amplitude $\bm{\alpha} = \tfrac{1}{\sqrt{2 \hbar }} (\bm{q}+i \bm{p})$, which is complex-normal distributed with mean $\tfrac{1}{\sqrt{2 \hbar}} (\bm{\bar{q}}+ i \bm{\bar{p}})$ and covariance matrix ${\bm{\Sigma}} \in \mathbb{C}^{2m \times 2m}$~\cite{picinbono1996second}.

\begin{center}
\begin{figure}[t!]
\includegraphics[width= 0.8\columnwidth]{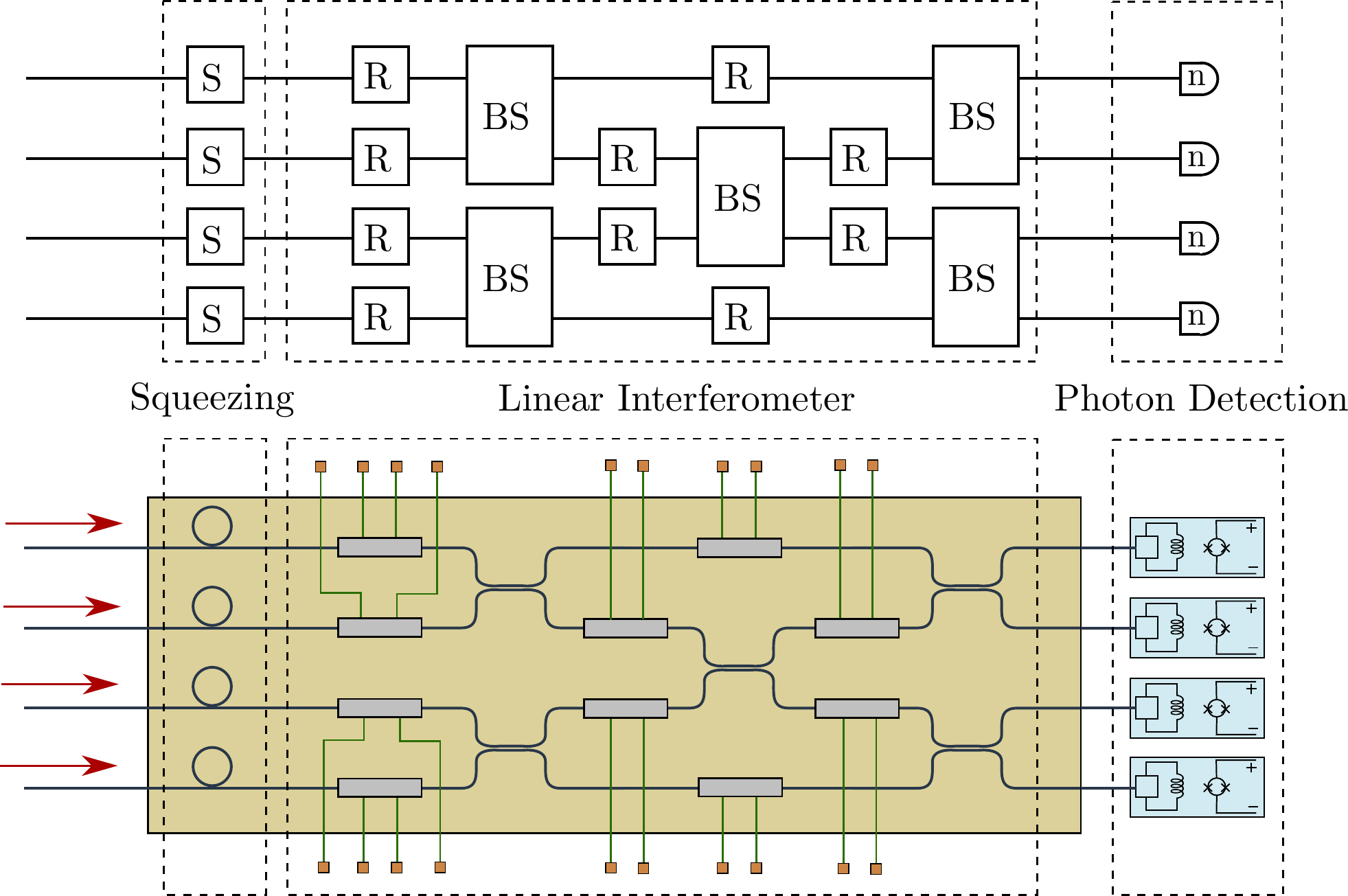}
\caption{A schematic illustration of a GBS circuit creating and measuring a Gaussian state with zero displacement. The top figure shows the sequence of gates that are used to create the Gaussian state: squeezing gates, followed by a linear interferometer that can be decomposed in terms of rotation (R) and beamsplitter (BS) gates. The state is then measured in the Fock basis, revealing the number of photons detected in each mode. The bottom figure is a schematic of a photonic chip realizing the same circuit. A bright pump laser enters the chip, which is used to generate squeezing in a neighbouring mode via nonlinear effects in a microring resonator. Rotation gates are implemented using tunable phase-shifters (grey boxes with electrical contacts). Beamsplitters are implemented through evanescent coupling of waveguides that are brought close to each other. They can be made tunable by extending them to a Mach-Zehnder interferometer, which is not shown in this figure. Finally, output light is measured using photon-number-resolving detectors, which can be implemented for example using superconducting transition edge sensors. 
}\label{Fig:GBS}
\end{figure}
\end{center}

GBS is a special-purpose model of photonic quantum computation where a multi-mode Gaussian state is prepared and then measured in the Fock basis. A general pure Gaussian state can be prepared from the vacuum by a sequence of single-mode squeezing, multimode linear interferometry~\cite{FR-serafini2017,clements2016optimal,de2018simple,reck1994experimental}, and single-mode displacements. In terms of the creation and annihilation operators $a_i$ and $a_i^\dagger$ on mode $i$, a squeezing gate is given by $S(r_i)=\exp[r_i(a_i^{\dagger 2}-a_i^2)/2]$, a displacement gate by $D(\alpha_i)=\exp(\alpha_i a_i^\dagger - \alpha_{i}^* a_i)$, and the linear interferometer transforms the operators as
\beq
\begin{pmatrix}
a_1 \\a_2\\\vdots \\a_m\end{pmatrix}\longrightarrow
\begin{pmatrix}
a_1' \\a_2'\\\vdots \\a_m'\end{pmatrix}
 = \bm{U}\begin{pmatrix}
a_1 \\a_2\\\vdots \\a_m\end{pmatrix},
\eeq
where $\bm{U}$ is a unitary matrix. Fock-basis measurements can be implemented with the use of photon-number-resolving (PNR) detectors~\cite{lita2008counting, hadfield2009single}. A GBS circuit is schematically illustrated in Fig.~\ref{Fig:GBS}.

\vspace{-1cm}

It was shown in Refs.~\cite{hamilton2017,kruse2019detailed} that for a Gaussian state with zero mean---which can be prepared using only squeezing followed by linear interferometry---the probability $\Pr(S)$ of observing an output $S=(s_1, s_2, \ldots, s_m)$, where $s_i$ denotes the number of photons detected in the $i$-th mode, is given by
\beq\label{Eq:pr_haf}
\Pr(S) = \frac{1}{\sqrt{\text{det}(\bm{Q})} } \frac{\text{Haf}(\bm{\mathcal{A}}_S)}{s_1!s_2!\cdots s_m!},
\eeq
where
\begin{align}
\bm{Q}&:=\bm{\Sigma} +\id/2,\label{Eq: Q}\\
\bm{\mathcal{A}} &:= \bm{X} \left(\id- \bm{Q}^{-1}\right)\label{Eq:A_matrix},\\
\bm{X} &:=  \left[\begin{smallmatrix}
	0 &  \id \\
	\id & 0  
\end{smallmatrix} \right].
\end{align}
Note that both $\bm{Q}$ and $\bm{\mathcal{A}}$ are completely determined by the covariance matrix $\bm{\Sigma}$ of the Gaussian state. The submatrix $\bm{\mathcal{A}}_S$ is uniquely specified by the output pattern $S$ of detected photons. It is constructed by repeating columns and rows $(i,i+m)$ of $\bm{\mathcal{A}}$ as follows: if $s_i=0$, the rows and columns $i$ and $i+m$ are deleted from $\bm{\mathcal{A}}$ and, if $s_i>0$, the corresponding rows and columns are repeated $s_i$ times.

The matrix function ${\rm Haf}(\cdot)$ is the 
\emph{hafnian}~\cite{caianiello1953quantum, bjorklund2018faster, barvinok2016combinatorics}, defined as
\begin{equation}
{\rm Haf}(\bm{\mathcal{A}}) = \sum_{\pi\in {\rm PMP}} \prod_{(i,j)\in \pi} \mathcal{A}_{ij},
\end{equation}
where $\mathcal{A}_{ij}$ are the entries of $\bm{\mathcal{A}}$ and $\rm PMP$ is the set of perfect matching permutations. The hafnian is a generalization of the permanent, in the sense that for any matrix $\bm{C}$ it holds that 
\beq\label{Eq:PerHaf}
\text{Haf}\begin{pmatrix}
0 & \bm{C}\\
\bm{C}^T & 0
\end{pmatrix}=\text{Per}(\bm{C}).
\eeq 
Computing the hafnian is a \#P-hard problem, a fact that has been leveraged to argue that, unless the polynomial hierarchy collapses to third level, it is not possible to efficiently simulate GBS using classical computers~\cite{aaronson2013, hamilton2017}. Indeed, state-of-the-art classical simulation algorithms for GBS require exponential time in a general setting~\cite{quesada2018gaussian,gupt2018classical,quesada2019classical,wu2019speedup}. 

When the Gaussian state is pure, the matrix $\bm{\mathcal{A}}$ can be written as $\bm{\mathcal{A}}=\bm{A}\oplus \bm{A}^*$, with $\bm{A}$ an $m\times m$ symmetric matrix. In this case, the output probability distribution is given by
\beq\label{Eq:pr_haf^2}
\Pr(S) = \frac{1}{\sqrt{\text{det}(\bm{Q})} } \frac{|\text{Haf}(\bm{A}_S)|^2}{s_1!s_2!\cdots s_m!},
\eeq
where the submatrix is defined with respect to rows and columns $i$, not $(i, i+m)$.

In certain situations, it is only relevant to determine in which modes at least one photon is detected. Threshold detectors have precisely this property: they ``click" whenever one or more photons are observed. It was shown in Ref.~\cite{quesada2018gaussian} that the GBS output probability distribution with threshold detectors is given by 
\beq
\Pr(\mathcal{S}) = \frac{\text{Tor}(\bm{O}_\mathcal{S})}{ \sqrt{\det(\bm{Q})}},
\eeq 
where $\bm{O} = \id - \bm{Q}^{-1}$, and $\text{Tor} (\cdot)$ is the Torontonian function~\cite{quesada2018gaussian}. In this case $\mathcal{S}=(s_1, s_2, \ldots, s_m)$ with $s_i=0,1$ and $s_i=1$ signals a click in the detector.

In the most general case, Gaussian states have non-zero mean, i.e., they require displacements to be prepared. To describe the resulting distribution, we define $\bm{\bar \alpha}' := (\bm{\bar \alpha},\bm{\bar \alpha^*})^T$, $\bm{\gamma}  := \bm{Q}^{-1} \bm{\bar \alpha}'$, and the matrix $\bm{\mathcal{A}'}$ with entries
\beq
\mathcal{A}'_{ij} = \begin{cases}
\mathcal{A}_{ij} &\text{ if } i\neq j,\\
\gamma_{i} &\text{ if } i=j,
\end{cases}
\eeq
where $\gamma_{i}$ is the $i$-th entry of $\bm{\gamma}$. As shown in Refs.~\cite{bjorklund2018faster,quesada2019franck,quesada2019simulating}, the output probabilities are given by
\begin{align}\label{Eq: lhaf}
\Pr(S)  = \frac{\exp\left(-\tfrac{1}{2} \bm{\bar \alpha}'^\dagger \bm{Q}^{-1} \bm{\bar \alpha}' \right)}{ \sqrt{\text{det}(\bm{Q})} s_1!s_2!\cdots s_m!}  \text{lhaf}(\bm{\mathcal{A}'}_S),
\end{align}
where lhaf is the \emph{loop hafnian} introduced in Ref.~\cite{bjorklund2018faster}. 

Practical implementations of quantum devices will inevitably be subject to experimental imperfections.
For near-term devices without error correction, it is important to study how quantum algorithms
perform in the presence of noise. A dominant type of noise in quantum-optical devices is photon loss,
arising from sources such as optical coupling~\cite{wang2019integrated,wang2018toward}. Loss on mode $i$ with creation operator $a_{i}$ can be modeled as a channel that couples to another mode in the vacuum with creation operator $b_{i}$,
resulting in the transformation
\begin{equation}
a_{i} \rightarrow a_{i} \sqrt{1 - L} + b_{i} \sqrt{L}
\end{equation}
followed by a tracing out of the ancilla mode. Here, $L \in [0, 1]$ is the loss parameter
describing the proportion of photons lost. One simple approach to simulating noise is to model a system with identical loss channels of equal strength acting on each mode.

\subsection{Programming a GBS device}\label{Sec:Programming}

A quantum computer is typically programmed by specifying a sequence of elementary gates. Different algorithms vary in the number, type, and order of these gates. Variational algorithms operate in a conceptually different manner: a sequence of parametrized gates are pre-specified and fixed by the algorithm, and the description of the program consists of listing the parameters for each gate~\cite{peruzzo2014variational,mcclean2016theory}. GBS can also be viewed in this manner: it is built from a sequence of predetermined Gaussian gates: squeezing, linear interferometry and displacement gates, whose parameters determine the Gaussian state to be sampled. 

In pure-state GBS without displacements, specifying gate parameters is equivalent to specifying the symmetric matrix $\bm{A}$ that characterizes the GBS probability distribution of Eq.~\eqref{Eq:pr_haf^2}. Employing the Takagi-Autonne decomposition~\cite{horn1990matrix, cariolaro2016, cariolaro2016reexamination}, we can write
\beq\label{Eq:Takagi}
\bm{A} = \bm{U} \text{diag}(\lambda_1, \lambda_2, \ldots, \lambda_m) \bm{U}^T,
\eeq   
where $\bm{U}$ is a unitary matrix. As shown in Refs.~\cite{bradler2018gaussian,jahangiri2019point}, $\bm{U}$ is precisely the unitary operation that specifies the linear interferometer of a GBS device. The values $0\leq \lambda_i < 1$ uniquely determine the squeezing parameters $r_i$ via the relation $\tanh (r_i) = \lambda_i$, as well as the mean photon number $\bar{n}$ of the distribution from the expression
\beq\label{Eq: mean_photon}
\bar{n} = \sum_{i=1}^M \frac{\lambda_i^2}{1-\lambda_i^2}.
\eeq

It is possible to encode an arbitrary symmetric matrix $\bm{A}$ into a GBS device by rescaling the matrix
with a parameter $c>0$ such that $c\bm{A}$ satisfies $0\leq \lambda_i < 1$ as in the above decomposition.
The parameter $c$ controls the values $\lambda_i$ and therefore also the squeezing parameters $r_i$ and the mean photon number $\bar{n}$. Overall, a GBS device can be programmed as follows:
\begin{enumerate}
\item Compute the Takagi-Autonne decomposition of $\bm{A}$ to determine the unitary $\bm{U}$ and the values $\lambda_1, \lambda_2, \ldots, \lambda_m$ as in Eq.~\eqref{Eq:Takagi}.
\item Program the linear interferometer according to the unitary $\bm{U}$. 
\item Solve for the constant $c>0$ such that $\bar{n} = \sum_{i=1}^M \frac{(c\lambda_i)^2}{1-(c\lambda_i)^2}$.
\item Program the squeezing parameter $r_i$ of the squeezing gate $S(r_i)$ acting on the $i$-th mode as $r_i=\tanh^{-1}(c\lambda_i)$.
\end{enumerate}

The GBS device then samples from the distribution
\beq\label{Eq: GBS_symm}
\Pr(S) \propto c^{k} \frac{|\text{Haf}(\bm{A}_S)|^2}{s_1!\ldots s_m!},
\eeq
with $k = \sum_{i}s_{i}$. Alternatively, of course, the device can be programmed by directly specifying the Gaussian state to be sampled. However, in most of the known applications, it is more helpful to think in terms of encoding symmetric matrices into the device.

Arguably, more applications of GBS have been developed compared to other Boson Sampling models because a wider class of problems can be encoded into GBS. As we have just learnt, a GBS device can be programmed according to any symmetric matrix. One example structure that can be described in terms of a symmetric matrix is a graph. Formally, a graph $G=(V, E)$ is composed of a set of nodes $V$ and a set of edges $E$. An undirected graph can be represented through its
symmetric adjacency matrix $\bm{A}$ with entries defined as
\beq\label{Eq:Adj}
A_{ij}=\begin{cases} &w_{ij} \text{    if } (i,j)\in E\\
&0 \text{    otherwise}, \end{cases}
\eeq
where an edge $(i,j)$ denotes a connection between nodes $i$ and $j$ with weight $w_{ij}$ (simple unweighted graphs have $w_{ij}=1$). Therefore, a GBS device can be programmed according to any undirected graph. We discuss in the
upcoming sections how encoding a graph can be used to solve graph-based problems.
This applies also to GBS with threshold detectors because the programming of squeezing parameters and linear interferometer is identical; only the detectors change. 

\section{The Strawberry Fields applications layer}

\begin{center}
\begin{figure}[t!]
\includegraphics[width= 0.83\columnwidth]{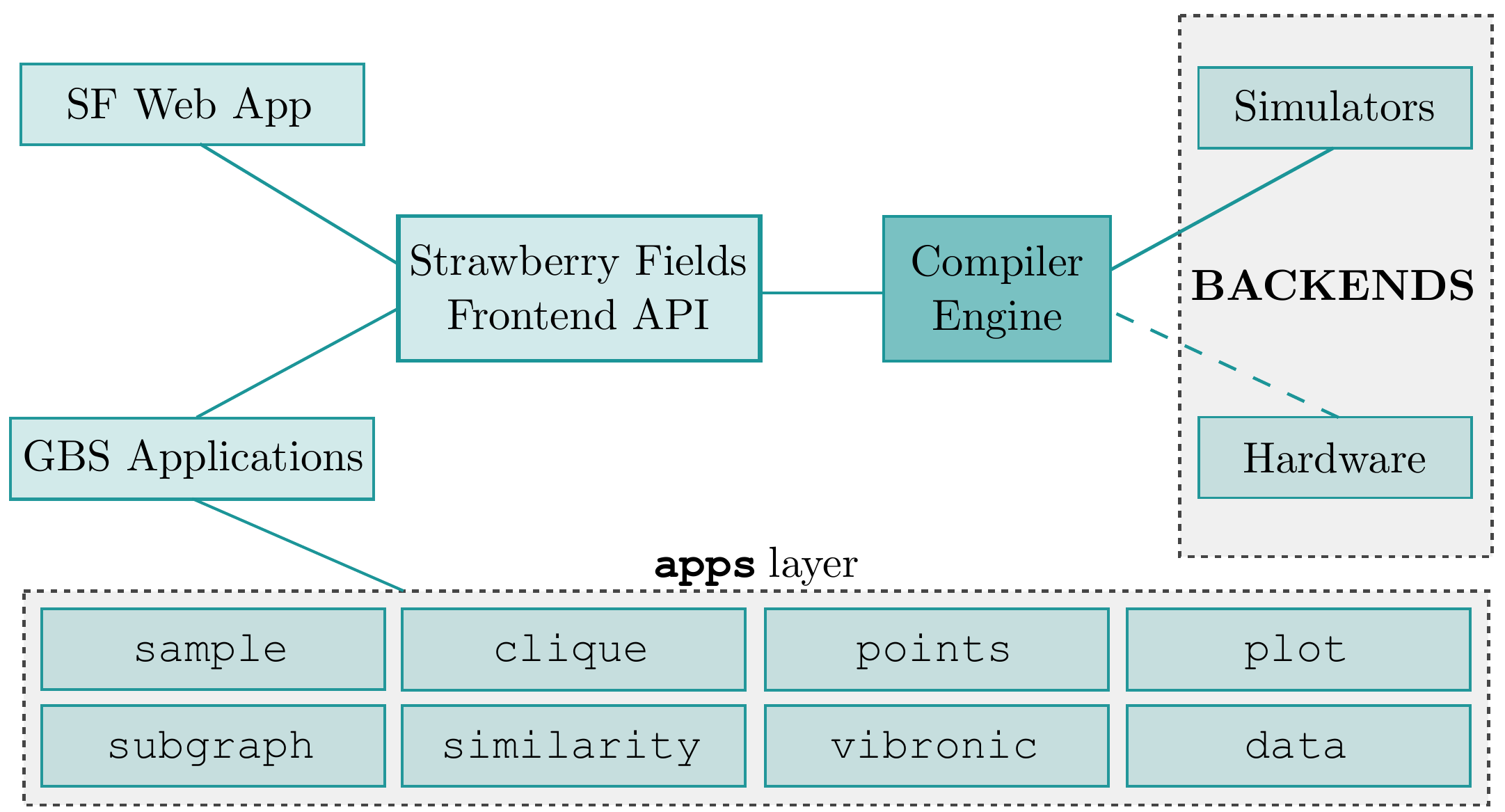}
\caption{A schematic depiction of the Strawberry Fields library for photonic quantum computing.
The Strawberry Fields frontend API consists of several functionalities, including operations, 
decompositions, and feed-forward capability. Quantum programs designed using the front-end 
are passed to the compiler engine, which converts them into elementary commands executable on different
back-ends. Simulation-based backends are currently available, with hardware planned to come online
in the future. The \python{apps} layer is a user-facing component that implements GBS
algorithms. It consists of application-specific modules for dense subgraph identification, maximum clique, graph similarity, point processes, and vibronic spectra. The remaining modules provide support for sampling, visualization, and for accessing GBS data.
}\label{Fig:SF}
\end{figure}
\end{center}

Strawberry Fields is an open-source Python library for photonic quantum computing~\cite{strawberryfields}. It consists of several components: front-end modules for construction and compilation of quantum programs, back-end components for running programs on simulators and hardware, and a web application. The GBS applications layer is a new front-end component that focuses on providing all the tools required to implement GBS algorithms. A schematic architecture diagram for the Strawberry Fields library, including the applications layer, is shown in Fig.~\ref{Fig:SF}. 

The applications layer is organized in modules, most dedicated to a specific GBS application: dense subgraph identification, maximum clique, graph similarity, point processes, and vibronic spectra. Auxiliary modules are also available that provide support for sampling, visualization, and access to pre-generated data. Users interested in a specific application can therefore work almost exclusively within the designated module. 
\vspace{-1cm}

Strawberry Fields can be installed following the instructions in the online documentation. For \bash{pip} users, this is as straightforward as running the command \bash{pip install strawberryfields}. Within the Strawberry Fields library, the applications layer is referred to as \python{apps}, and its modules can be accessed using standard Python import commands. A supporting set of tutorials for each application is available in the Strawberry Fields documentation, which also provides links to download the tutorials as executable scripts and Jupyter Notebooks~\cite{kluyver2016jupyter}.

A main functionality of the applications layer is to provide users with the capability to generate samples from a GBS device. This is performed with the \python{sample} module. As discussed in the previous section, a GBS device can be programmed by specifying a symmetric matrix and a mean photon number. To generate samples, a user imports the module, defines the matrix and mean photon number, and calls the \python{sample()} function. This is shown in the example code below for generating GBS samples from a random $10\times 10$ symmetric matrix:
\begin{customcode}
>>> from strawberryfields.apps import sample
>>> import numpy as np
>>> modes = 10  # number of modes
>>> n_mean = 6  # mean photon number
>>> n_samples = 2  # number of samples
>>> A = np.random.normal(0, 1, (modes, modes))  # random Gaussian matrix
>>> A = A + A.T  # symmetrizes matrix
>>> samples = sample.sample(A, n_mean, n_samples, threshold=False, loss=0.1)
\end{customcode}
The \python{sample()} function interfaces with the \python{gaussian} backend of Strawberry Fields, which in turn passes to The Walrus library~\cite{walrus} that hosts optimized sampling algorithms. Optionally, the choice of threshold or PNR detection can be input, defaulting to threshold. The proportion of photons lost in the simulation due to equal-strength noise channels applied to each mode can also be set, defaulting to zero loss. In the example code the loss was set to 10\%.

The \python{sample} module also provides users with the ability to postselect samples based on a range of desired photon numbers. Continuing from the code above, it is possible to postselect samples with photon numbers between four and ten:
\begin{customcode}
>>> n_min = 4 
>>> n_max = 10
>>> samples = sample.postselect(samples, n_min, n_max)
\end{customcode}

The applications layer includes a \python{data} module containing pre-generated samples from a variety of problems. Samples can be accessed by loading the corresponding class. For example, the class \python{PHat} contains 50,000 samples from the 300-node \bash{p_hat300-1} graph of the DIMACS dataset~\cite{johnson1993cliques, gendreau1993solving}, generated with a mean photon number of $\bar{n}=10$. Accessing the samples and other information can be done by interacting with \python{PHat}:
\begin{customcode}
>>> from strawberryfields.apps import data
>>> phat = data.PHat()  # loads data
>>> phat[:100]  # first 100 samples
>>> A = phat.adj  # adjacency matrix
>>> phat.n_mean  # mean photon number
10
\end{customcode}

Two important packages used within the applications layer are NetworkX~\cite{hagberg2008exploring}
and Plotly~\cite{plotly}, which are also open source and \bash{pip} installable. NetworkX
is used to provide a representation of graphs using the \python{Graph} class, along with a wide
range of tools for graph manipulation and analysis. An adjacency matrix \python{A} can be converted into a
NetworkX \python{Graph} with:
\begin{customcode}
>>> import networkx as nx
>>> g = nx.Graph(A)
\end{customcode}
The Plotly package is used in the \python{plot} module of the
applications layer. Each function returns a figure which can be
automatically plotted in Python's interactive mode:
\begin{customcode}
>>> from strawberryfields.apps import plot
>>> plot.graph(g)
\end{customcode}

We have thus far introduced the basic principles underlying GBS devices and have given an overview of the Strawberry Fields applications layer. In the next section, we bring these concepts together, covering each application of GBS in detail, while describing how the applications layer can be used to implement the corresponding GBS algorithms.

\section{Applications of Gaussian Boson Sampling}
In this section, we describe in detail all currently known GBS algorithms. Each application has a dedicated subsection, meaning that they are covered in a self-contained manner. In each case, we provide an overview of the problem, a description of the corresponding GBS algorithm, and example code outlining how to use the applications layer to implement the algorithms. This section therefore serves jointly as a short review of GBS algorithms and as an introduction to the use of the applications layer. More detailed documentation and tutorials for the applications layer can be found online.

As is common for applications running on near-term devices, most of the GBS algorithms we describe are \emph{heuristic} approaches to solving problems. They constitute novel methods that leverage the unique properties of GBS devices. In some cases, GBS serves as an accelerator for classical algorithms; in others, as a method for building statistical models. GBS devices can also be used to compute objects, such as histograms or vectors, that reveal desired statistical properties of the encoded problems.

The algorithms we describe constitute the current state of the art in applications of GBS, which are the result of only a few years of research. They reflect the nascence of a growing field, which may be additionally fueled by rapid progress in experimental implementations. Indeed, one of the goals of the applications layer is to provide researchers with tools to enable further innovation in the development of algorithms for near-term photonic quantum computers.

\subsection{Dense subgraph identification}\label{Sec:subgraphs}

Graphs can be used to model a wide variety of concepts including social networks~\cite{angel2012dense},
websites~\cite{kumar1999trawling}, financial markets~\cite{arora2011computational}, and
biological networks~\cite{fratkin2006motifcut,saha2010dense}. A common problem of interest is to
identify dense subgraphs, i.e., subgraphs that contain a large number of connections between their nodes. Dense subgraphs represent subsets of nodes that are highly connected, which may correspond for example
to communities in social networks or to mutually influential proteins in a biological network. In a nutshell, dense subgraphs are the highly correlated regions of graphs, and finding dense subgraphs is relevant whenever identifying such correlations is important.

The dense subgraph problem can be stated formally as follows. Let $\mathcal{S}=(s_1, s_2, \ldots, s_m)$ denote a subset of the $m$ nodes, with 
$s_i=1$ if the node is part of the subset, and $s_i=0$ otherwise. The \emph{induced subgraph} $G[\mathcal{S}]$ is the graph made from nodes in $\mathcal{S}$ and all the edges whose endpoints are in $\mathcal{S}$. Let $d(\mathcal{S})=2|E(\mathcal{S})|/[|\mathcal{S}|(|\mathcal{S}| - 1)]$ be the \emph{density} of the subgraph, where $|E(\mathcal{S})|$ is the number of edges and $|\mathcal{S}|$ is the number of nodes. A complete graph has $|\mathcal{S}|(|\mathcal{S}| - 1)/2$ edges and unit density.

The dense subgraph problem is the task of finding the densest $k$-node subgraph, i.e.,
\begin{equation}
{\rm argmax}_{\mathcal{S}}\{ d(\mathcal{S}) \, \,  : \, \, |\mathcal{S}| = k\}.
\end{equation}
The optimization can further proceed over a range of $k$. More generally, it is also of interest to identify multiple regions of high density in a graph. This involves keeping track of many subgraphs, not just the densest.

\subsubsection{GBS algorithm for dense subgraph identification}
We describe an algorithm that uses GBS to identify dense subgraphs. A graph can be naturally encoded into GBS through its symmetric adjacency matrix $\bm{A}$, using the embedding discussed in Sec.~\ref{Sec:Programming}. From Eq.~\eqref{Eq: GBS_symm},
the probability of a sample $S = (s_{1}, s_{2}, \ldots, s_{m})$
is proportional to the square of the hafnian of the submatrix
$\bm{A}_S$. If the sample $S$ is
composed entirely of zero and single photon counts, we can identify it with a subset of nodes $\mathcal{S}$, in which case $\bm{A}_{S}$ is precisely the adjacency matrix of the induced subgraph $G[\mathcal{S}]$.

More generally, the sample $S$ may contain
multi-photon events which describe an extended induced
subgraph $G[S]$~\cite{schuld2019quantum}. Here, the sample $S$ is again used to pick out nodes from $G$, but each node $i$ is replicated $s_{i}$ times along with all of its connections. The important lesson is that we can uniquely interpret GBS outputs as subgraphs, which are sampled according to a probability that depends on the hafnian of their corresponding adjacency matrix.

\begin{center}
\begin{figure}[t!]
\includegraphics[width= \columnwidth]{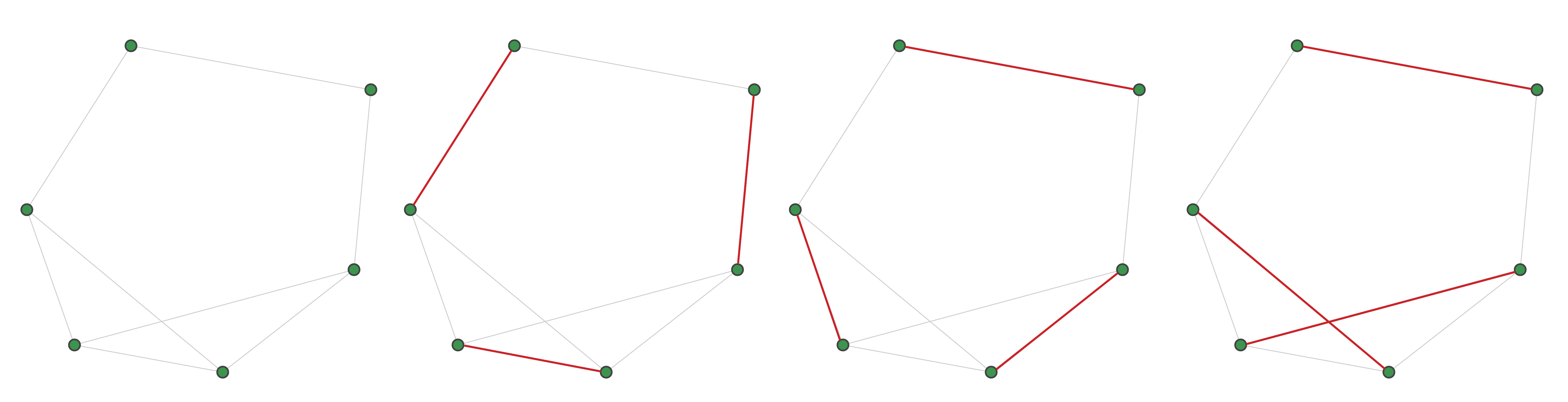}
\caption{All the perfect matchings in a graph of six nodes. The graph is shown on the left. The three perfect matchings in the graph are shown on the right by highlighting the subset of edges that constitute the perfect matching. Note that in each case, every node is paired with a unique partner node. The hafnian of the graph's adjacency matrix is equal to three.}\label{Fig:PMs}
\end{figure}
\end{center}

The hafnian of an adjacency matrix $\bm{A}$ is equal to the number of \emph{perfect matchings} in the graph~\cite{barvinok2016combinatorics}. A perfect matching is a subset of the edges such that each node in the graph is connected by an edge with another unique node. This way, every node is perfectly matched with a partner, as illustrated in Fig.~\ref{Fig:PMs}. In general, there may be many possible perfect matchings in a graph $G$. We denote this number by $N_{\rm PM}(G)$. Counting the number of perfect matchings in a graph, which is a special case of computing the hafnian, is also a \#P-hard problem. The probability of sampling subgraphs from GBS, Eq.~\eqref{Eq: GBS_symm}, can then be written as
\begin{equation}\label{Eq:GBSGraph}
\Pr(G[S]) \propto \frac{[N_{\rm PM}(G[S])]^2}{s_1!s_2!\cdots s_m!}.
\end{equation}
In other words, samples from a GBS device represent subgraphs and subgraphs that have a large number of perfect matchings are sampled with high probability.

In other words, we can program a GBS device such that, with high probability, it samples subgraphs with a large number of perfect matchings.
\vspace{-1cm}

It was discussed in~\cite{arrazola2018using} that the number of perfect matchings in a graph $G$
correlates with the density of the graph $d(G)$. More concretely, the number of
perfect matchings is upper bounded by a monotonically increasing function of the number of edges in a graph~\cite{aaghabali2015upper}. This fact implies a correlation between hafnian and density: graphs with a large hafnian must also have high density. Therefore, a GBS device samples dense subgraphs with high probability, which is useful if the goal is to identify such dense subgraphs.

This insight can be employed to create a GBS algorithm for dense subgraph identification. It is a hybrid algorithm based on a classical strategy due to Charikar~\cite{charikar2000greedy} that resizes subgraphs by adding or removing nodes based on node degree (number of incident edges to a given node). We outline the algorithm below:

\begin{center}
\textit{Algorithm}
\end{center}
\begin{enumerate}
\item Encode the graph $G$ into the GBS device through its adjacency matrix $\bm{A}$ and generate $N$ samples from a distribution with mean photon number $\bar{n}$.
\item Select a range $[k_\text{min}, k_\text{max}]$ of subgraph sizes.
\item Let $k$ denote the size of a subgraph. For each of the $N$ sampled subgraphs, if $k>k_\text{min}$, remove the node with the lowest degree relative to the subgraph, breaking ties uniformly at random. Continue removing nodes this way until the resized subgraph has $k_\text{min}$ nodes. 
\item Similarly, for each sampled subgraph, if $k<k_\text{max}$, add the node with the highest degree relative to the subgraph, breaking ties uniformly at random. Continue adding nodes until the resized subgraph has $k_\text{max}$ nodes. 
\item Output the densest subgraphs found by the algorithm for each size in the range $[k_\text{min}, k_\text{max}]$.
\end{enumerate}

The role of GBS is to ``seed'' the dense subgraph search by providing starting points that are already dense subgraphs with high probability, thus acting as an accelerator to the algorithm. Charikar resizing then proceeds normally, allowing a sweep through subgraphs of varying size. The choice of mean photon number $\bar{n}$ is important since it determines the typical sample size obtained from GBS, which should ideally be matched with the size where we expect the densest subgraphs to be found. In Refs.~\cite{arrazola2018using, arrazola2018quantum}, numerical evidence was presented showing that the performance of a wide range of classical algorithms can indeed be improved using GBS as a seed.

\subsubsection{Using the GBS applications layer}
\vspace{-1cm}

\begin{center}
\begin{figure}[t!]
\includegraphics[width= 0.4\columnwidth]{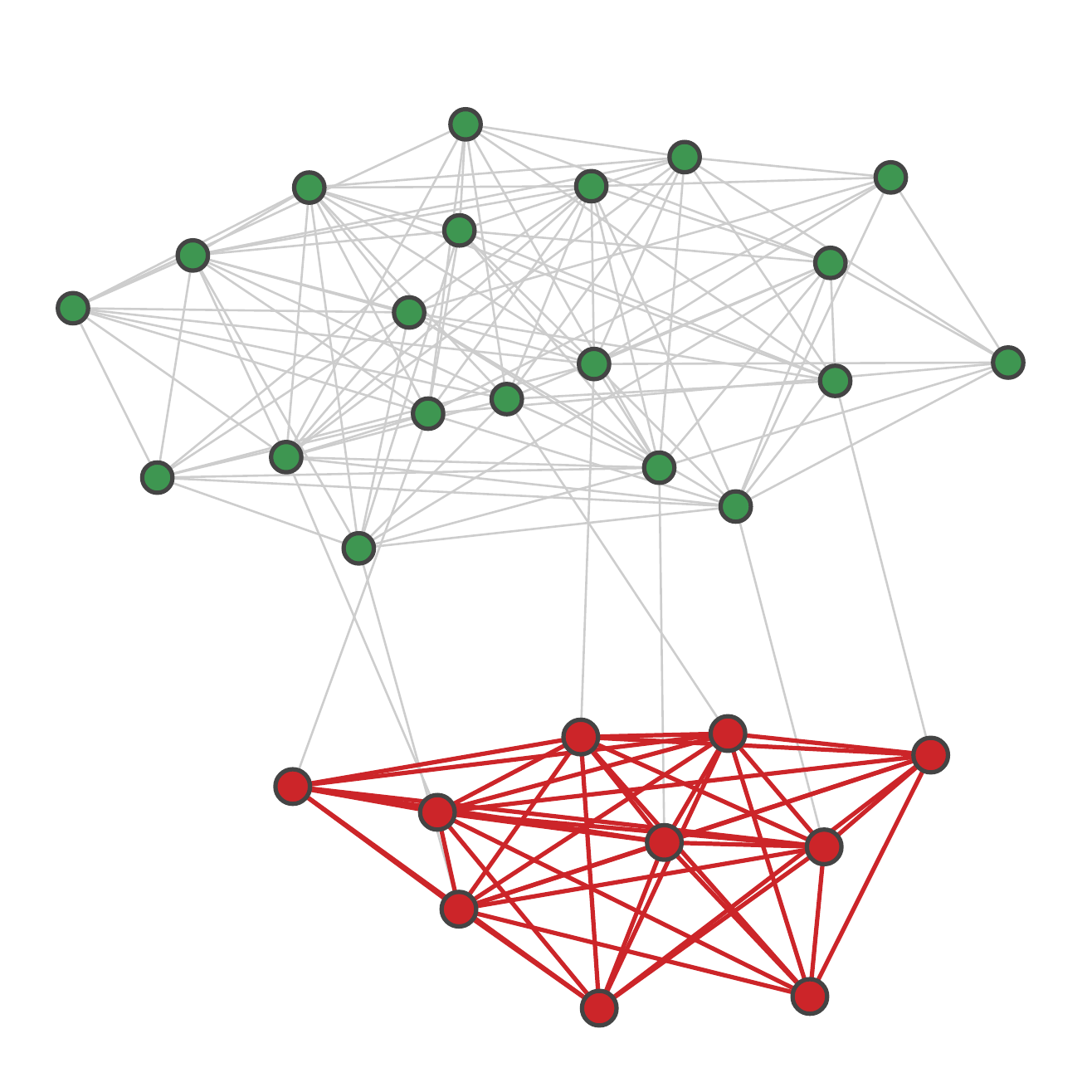}
\caption{A graph generated by joining two Erd\H{o}s-R\'enyi random graphs:
the first graph of 20 nodes (shown on top in green) is created with edge probability of 0.5. The second planted graph of 10 nodes (shown below in red) is generated with edge probability of 0.875. The two graphs are joined by selecting
8 nodes at random from
both graphs and adding an edge between them. This figure was produced using the \python{plot} module of the applications layer. }\label{Fig:Planted}
\end{figure}
\end{center}

The \python{subgraph} module of the applications layer is dedicated to dense subgraph identification. It provides functions for resizing graphs based on node degree and for searching dense subgraphs using the GBS algorithm described above. Here we focus on a 30-node graph containing a planted dense 10-node graph, which is shown in
Fig.~\ref{Fig:Planted}~\cite{arrazola2018using}. The planted subgraph has the property that, even though it is the densest of its size, its nodes have a lower average degree than those in the rest of the graph. This poses a challenge for
algorithms based on node degree. To investigate this, we can
use the \python{resize()} function in the \python{subgraph} module, which implements the algorithm for subgraph resizing. The \python{data} module contains the class \python{Planted}, which includes 50,000 samples from this graph with a mean photon number of $\bar{n}=8$.

\begin{customcode}
>>> from strawberryfields.apps import data, subgraph
>>> import networkx as nx
>>> g = nx.Graph(data.Planted().adj)  # load graph
>>> s = range(30) # starting subgraph is the entire graph
>>> r = subgraph.resize(s, g, min_size=1, max_size=29)  # resize subgraph
>>> r[10]  # output subgraph of size 10
[0, 6, 8, 9, 10, 12, 13, 14, 15, 17]
\end{customcode}

In this case, the resizing fails to identify the dense planted subgraph (which consists of the last ten nodes) since its nodes have low degree and therefore are removed in the first steps of resizing. In fact, the output graph does not have any overlap with the planted subgraph. This difficulty can be addressed by using GBS samples as starting points. The GBS algorithm for dense subgraph identification is provided by the
\python{search()} function, with the input subgraphs sampled from GBS. The \python{search()} function applies \python{resize()} to each subgraph and
keeps track of the number of densest subgraphs identified for each size. It
returns a dictionary over sizes, with each value a list of densest subgraphs
for a given size, described by their density and nodes. Its implementation using the applications layer is shown below:

\begin{customcode}
>>> from strawberryfields.apps import sample
>>> s = data.Planted()  # load samples from data module
>>> s = sample.postselect(s, 16, 30)  # postselect sample sizes
>>> s = sample.to_subgraphs(s, g)  # convert samples to subgraphs
>>> k_min = 8  # smallest subgraph size
>>> k_max = 16  # largest subgraph size
>>> r = subgraph.search(s, g, k_min, k_max)  # implement search algorithm
>>> r[10][0]  # densest subgraph of size 10
[(0.9333333333333333, [20, 21, 22, 23, 24, 25, 26, 27, 28, 29])]
\end{customcode}

We see from the code above that the algorithm was able to find the planted subgraph with density $0.933$. It is also often useful to identify many high-density subgraphs in a range of different sizes. Continuing from the code above, we can extract the densest subgraphs of size 8 and 12:

\begin{customcode}
>>> r[8][0]  # densest subgraph of size 8
[(1.0, [21, 22, 24, 25, 26, 27, 28, 29])]
>>> r[12][0]  # densest subgraph of size 12
[(0.696969696969697, [0, 2, 3, 5, 6, 8, 9, 10, 14, 16, 17, 18])]
\end{customcode}

Depending on their size,
the densest subgraphs belong to different regions of the graph: dense subgraphs of less than
ten nodes are contained within the planted subgraph, whereas larger dense subgraphs appear
outside of the planted subgraph. Smaller dense subgraphs can be cliques, characterized by
having maximum density of 1, while larger subgraphs are less dense.

\subsection{Maximum clique}\label{Sec:MaxClique}

A clique is a subgraph that contains all possible edges between its nodes---it is a complete graph with unit density. The maximum clique problem, or max clique for short, is the task of finding the largest clique in a graph~\cite{bomze1999maximum, wu2015review}. Mathematically, it can be stated as
\begin{equation}
{\rm argmax}_{\mathcal{S}} \{|\mathcal{S}| \,\, : \,\, d(\mathcal{S}) = 1\}.
\end{equation}

Max clique is NP-hard and its decision version is one the the most widely-studied NP-complete problems~\cite{karp1972reducibility}. Applications of max clique have been known for decades~\cite{pardalos1994maximum, bomze1999maximum}, and new applications continue to be discovered in a wide range of disciplines such as bioinformatics~\cite{malod2010maximum, ravetti2008identification}, social network analysis~\cite{balasundaram2011clique, pattillo2012clique}, finance~\cite{boginski2006mining}, flight scheduling~\cite{dorndorf2008modelling}, and telecommunications~\cite{balasundaram2006graph, jain2005impact}. 

\subsubsection{GBS algorithm for max clique}

Due to the importance of max clique, many algorithms have been developed to search for large cliques in graphs~\cite{wu2015review}. Among these, arguably the best-performing algorithms are based on local search techniques that identify small cliques and search for larger ones in their neighbourhood~\cite{pullan2006dynamic,pullan2006phased}. Local search algorithms operate as follows. In the first stage, given an input graph $G$ and a starting clique $\mathcal{C}$, the algorithm evaluates the set
$c_{0}(\mathcal{C})$ of nodes in the remainder of the graph that are
connected to all of the nodes in $\mathcal{C}$. A single node is picked from $c_{0}(\mathcal{C})$
and added to the clique, resulting in a larger clique $\mathcal{C}$ and a new set
$c_{0}(\mathcal{C})$. This process is repeated until no further growth can occur. Elements can be selected from $c_{0}(\mathcal{C})$ uniformly
at random or with probability proportional to node degree.

The local search phase works by evaluating the set $c_{1}(\mathcal{C})$ of nodes
in the remainder of the graph that are connected to \emph{all but one}
of the nodes in the clique $\mathcal{C}$. If this set is not empty, a node is selected and swapped
with the corresponding node in the clique. This allows the
algorithm to move to a new clique that could not have been reached during growth. Local search algorithms alternate between growth and swapping, typically until a dead end is reached and a new starting point is then selected. Some methods reported
in the literature can include more advanced elements such as adding penalties to commonly
encountered nodes to discourage their repeated selection and perturbing the clique after searching to explore a new area of the search space~\cite{wu2015review}.

\begin{center}
\begin{figure}[t!]
\includegraphics[width= \columnwidth]{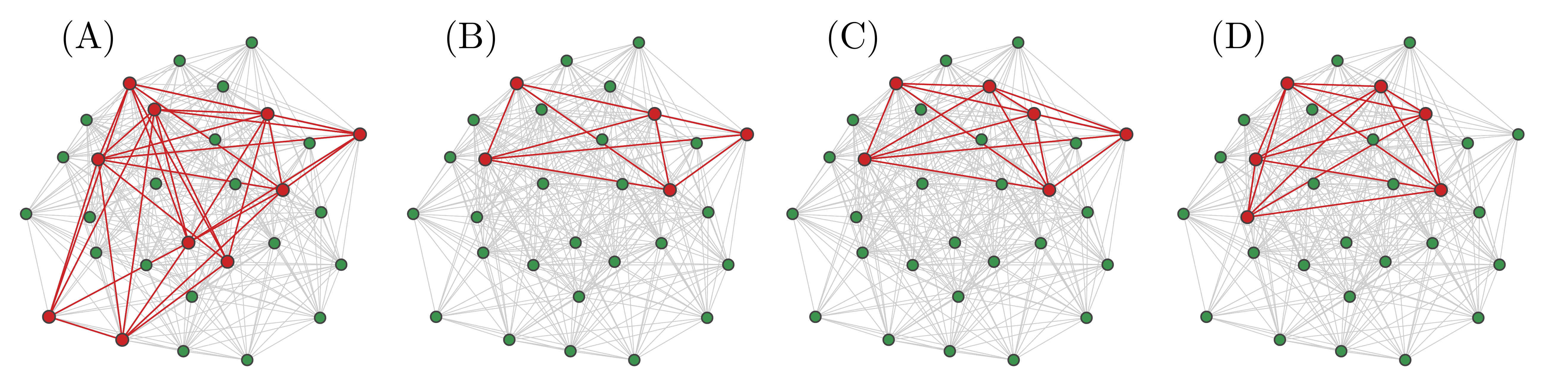}
\caption{An example of the GBS algorithm for max clique. The input is an Erd\H{o}s-R\'enyi graph of 30 nodes with edge probability 0.7. (A) The GBS device is programmed according to this graph and a sample is generated, in this case corresponding to a subgraph of 10 nodes, highlighted in red. (B) The subgraph is then shrunk by removing nodes of low degree until a clique $\mathcal{C}$ is formed, in this case of five nodes. (C) The clique is grown by computing the set $c_{0}(\mathcal{C})$ and adding a node in this set. (D) When no more growth is possible, a swap is performed by evaluating the set $c_{1}(\mathcal{C})$ and exchanging a node in $c_{1}(\mathcal{C})$ with its partner in the clique. This process of growth and swap is repeated until a dead end is reached, or a maximum number of iterations have occurred.}
\label{Fig:max_clique}
\end{figure}
\end{center}

Because finding large cliques is challenging, local search algorithms start by randomly selecting a single vertex in $G$. As discussed in section~\ref{Sec:subgraphs}, GBS devices can be programmed to sample dense subgraphs with high probability. Having unit density,
cliques are the most likely subgraphs to be sampled from GBS among subgraphs of the same size. Therefore, a way to improve local search algorithms is to sample dense subgraphs from GBS and use them as starting points for local search. Not all sampled subgraphs are guaranteed to be cliques, but a clique can be obtained from a dense subgraph by removing nodes with low degree until a clique is formed. Combining GBS with the local search algorithm results in a hybrid GBS algorithm that we describe below:

\begin{center}
\textit{Algorithm}
\end{center}
\begin{enumerate}
\item Encode the graph $G$ into the GBS device through its adjacency matrix $\bm{A}$ and generate $N$ samples from a distribution with mean photon number $\bar{n}$.
\item For each sampled subgraph, iteratively remove the node with the lowest degree relative to the subgraph, breaking ties uniformly at random, until a clique $\mathcal{C}$ is formed.
\item Evaluate the set $c_{0}(\mathcal{C})$. Grow the clique $\mathcal{C}$ by adding a node in $c_{0}(\mathcal{C})$ selected uniformly at random or based on node degree relative to the subgraph. Repeat until no further growth is possible.
\item Evaluate the set $c_{1}(\mathcal{C})$. Select a node in this set uniformly at random or based on node degree relative to the subgraph, and swap it with its partner in the clique. Go back to the growth step 3.
\item Repeat steps 3 and 4 until a dead end is found, i.e., until $c_{0}(\mathcal{C})$ and $c_{1}(\mathcal{C})$ are empty, or until a maximum number of iterations has been reached.
\item Output the largest clique found out of all $N$ samples.
\end{enumerate}
Note that it is possible to postselect GBS samples on a desired range of sizes before using them as seeds for local search. The GBS algorithm is illustrated in Fig.~\ref{Fig:max_clique}. 
\vspace{-1cm}

Max clique can be generalized to the \emph{weighted} maximum
clique problem, where the undirected graph has weights on its nodes and the goal is to
find the clique with maximum total weight~\cite{pelillo2009heuristics}. GBS can also be used to sample subgraphs from weighted-node graphs~\cite{banchi2019molecular}.
Consider a rescaling of the adjacency matrix $\bm{A}$ according to
\begin{equation}
\bm{A}' = \bm{\Omega} (\bm{D} - \bm{A}) \bm{\Omega},
\end{equation}
where $\bm{D}$ is the diagonal matrix of node degrees and $\bm{\Omega}$ is a tunable diagonal matrix. By
encoding $\bm{A}'$ into a GBS device and associating samples to subgraphs, the probability of sampling a subgraph $\mathcal{S}$ is
\begin{equation}
\mbox{Pr}(G[\mathcal{S}]) \propto [N_{\rm PM}(G[\mathcal{S}])]^2 \prod_{i \in \mathcal{S}}\Omega_{ii}^{2}.
\end{equation}
By setting $\bm{\Omega} = \bm{D}^{-\frac{1}{2}}$, $\bm{A}'$
becomes the normalized Laplacian~\cite{chung1997spectral}. For node-weighted graphs, control of $\bm{\Omega}$ can allow for a
trade-off between subgraph density and weight. One approach to introducing the node weights
$w_{i}$ is to set $\Omega_{ii} = 1 + \alpha w_{i}$. The value of
$\alpha$ determines a trade-off: for $\alpha \rightarrow 0$ the distribution becomes independent
of node weights, while a large $\alpha$ leads to subgraphs with high weight being the most likely.

In Ref.~\cite{banchi2019molecular}, the GBS algorithm for max clique on weighted graphs has been applied to the molecular docking problem. Here, the goal is to determine the docking configuration between a drug molecule and a target macromolecule, for example a protein in the human body. This task can be mapped to the weighted max clique problem~\cite{kuhl1984combinatorial}, and GBS devices can be used to improve the performance of classical algorithms. Indeed, it was reported in Ref.~\cite{banchi2019molecular} that employing GBS samples could increase the success rate of the algorithm roughly from 30\% to 70\%. 

\subsubsection{Using the GBS applications layer}
\begin{center}
\begin{figure}[t!]
\includegraphics[width= 0.31\columnwidth]{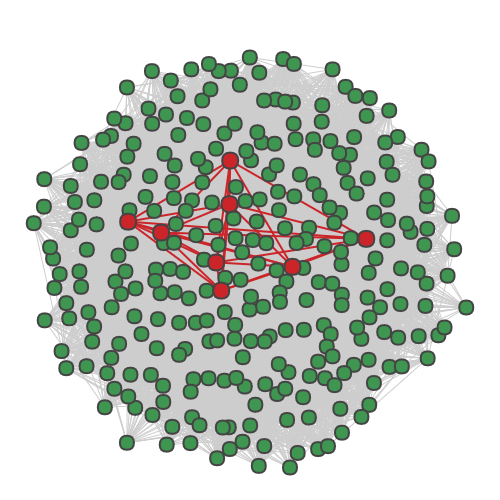}
\includegraphics[width= 0.31\columnwidth]{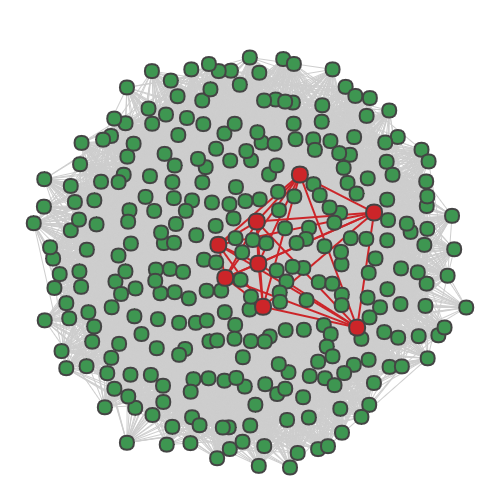}
\includegraphics[width= 0.31\columnwidth]{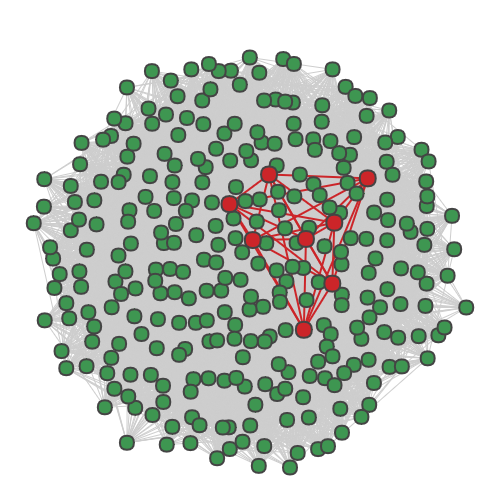}
\caption{The $300$-node \bash{p_hat300-1} random graph has maximum known clique
size of eight~\cite{gendreau1993solving}. This figure shows three different $8$-node cliques found by the GBS algorithm, which appear highlighted in red.}\label{Fig:PHat}
\end{figure}
\end{center}

The GBS applications layer contains the \python{clique} module, dedicated to the maximum clique problem. It includes functions for shrinking subgraphs to cliques and for implementing the local search algorithm. As a testbed graph for using GBS samples to help find the maximum clique, we use the $300$-node
\python{PHat} graph~\cite{johnson1996cliques}. This random graph, shown in Fig.~\ref{Fig:PHat},
has multiple maximum cliques of size $8$.
Pre-generated GBS samples are available in the \python{data}
module. 

The \python{clique} module contains two main functions: \python{shrink()} and \python{search()}. The first is used to shrink subgraphs to cliques, while the latter implements the local search algorithm. A full code for implementing the GBS hybrid algorithm using the applications layer is shown below: 
\begin{customcode}
>>> from strawberryfields.apps import clique, data, sample
>>> import networkx as nx
>>> p_hat = data.PHat()  # load samples 
>>> g = nx.Graph(p_hat.adj)  # create graph from adjacency matrix
>>> s = sample.postselect(p_hat, 16, 20)  # postselect samples
>>> s = sample.to_subgraphs(s, g)  # convert samples to subgraphs
>>> # shrink subgraphs to a list of cliques
>>> cliques = [clique.shrink(i, g) for i in s]
>>> # run local search for all cliques
>>> cliques = [clique.search(c, g, 10) for c in cliques]
\end{customcode}
The output of this code is a list of cliques found by the algorithm. The results can be analyzed further to determine properties of the cliques:

\begin{customcode}
>>> # sort cliques in decreasing size
>>> cliques = sorted(cliques, key=len, reverse=True)
>>> cliques[:3] # the three largest cliques
[[48, 53, 87, 152, 243, 273, 279, 295],
 [37, 78, 158, 207, 218, 239, 249, 267],
 [17, 48, 106, 148, 170, 196, 224, 234]]
>>> from strawberryfields.apps import plot
>>> p0 = plot.graph(g, cliques[0])  # See Fig. \ref{Fig:PHat} (left)
>>> p1 = plot.graph(g, cliques[1])  # See Fig. \ref{Fig:PHat} (middle)
>>> p2 = plot.graph(g, cliques[2])  # See Fig. \ref{Fig:PHat} (right)
\end{customcode}
Three cliques of size 8, which for this graph are known to be maximum cliques, are shown in Fig.~\ref{Fig:PHat}.
\vspace{-1cm}

\subsection{Graph similarity}
In the applications studied so far, a graph is encoded into a GBS device and properties of the resulting distribution are harnessed to identify optimal structures. The properties of the graph are therefore also transferred to the GBS distribution. In this section, we study how to analyze GBS samples obtained from different encoded graphs to obtain insights on the differences between their respective graphs.

Graph isomorphism is the problem of determining whether two graphs $G$ and $G'$ are isomorphic, i.e., whether they are identical up to a permutation of their nodes~\cite{read1977graph}. Mathematically, a graph isomorphism is a bijective mapping $f$ from the node set of $G$ to the node set of $G'$ such that an edge $(i,j)$ exists in $G$ if and only if the edge $(f(i), f(j))$ occurs in $G'$. Two graphs are isomorphic if such an isomorphism exists.  

More generally, graphs may be non-isomorphic yet
still share a similar structure. One approach to measure similarity is to introduce a \emph{feature map} that maps each graph
to a feature vector in a high-dimensional space~\cite{gartner2003graph,vishwanathan2010graph,shervashidze2011weisfeiler, grover2016node2vec,zhang2018network,goyal2018graph,ghosh2018journey}. Similarity between graphs can then be measured with a \emph{kernel,} which quantifies the distance between the vectors in the embedded space, for example using
the inner product. Since many objects from networks to molecules can be modeled as graphs, graph kernels constitute an approach to analyze and classify a wide variety of data. The challenge remains to build better mappings and kernels that reliably capture the graph properties that are relevant in specific applications.

\subsubsection{Graph similarity with GBS}

It was shown in Ref.~\cite{bradler2018graph} that two graphs are isomorphic if and only if their probability distributions from GBS are equal up to permutation. Hence, finding a single probability for
which the distributions differ is enough to conclude that two graphs are not isomorphic. 
However, it is not clear a priori which of the exponentially many probabilities should differ, and estimating individual probabilities by sampling from the GBS distribution of a graph is experimentally prohibitive since each probability is typically exponentially small.

Beyond isomorphism, GBS probability distributions can be used to measure the similarity between graphs, as we now detail. One approach to defining an experimentally accessible distribution is to \emph{coarse-grain} samples into more commonly occurring events. A
coarse-graining method introduced in Ref.~\cite{bradler2018graph} consists of combining all samples that are equivalent
under permutation. The set of all such samples is called an orbit. It can
be represented as a sorting of a sample in non-increasing order with the trailing zeros
removed. The orbit determines a specific pattern of detected photons and samples are coarse-grained by combining all outputs that belong to the same orbit. For example, a sample $S = (0, 2, 3, 0, 0, 0, 1, 2, 0)$ belongs to the orbit $(3, 2, 2, 1)$.

\begin{center}
\begin{figure}[t!]
\includegraphics[width= 0.7\columnwidth]{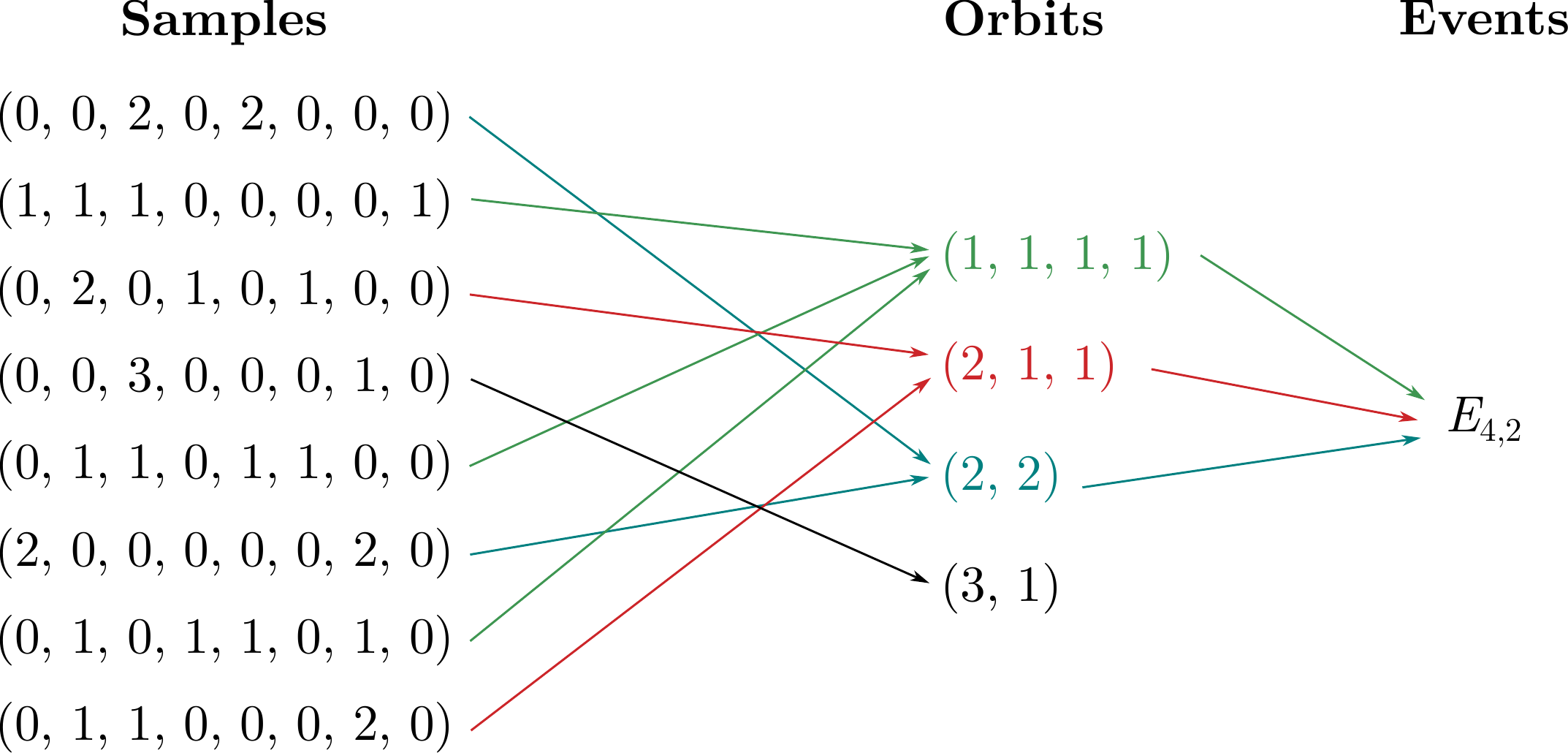}
\caption{An illustration of how samples are coarse-grained into orbits and events. In this example, we consider eight samples containing $k=4$ photons. They can be coarse-grained into three different orbits: $(1,1,1,1)$, $(2,1,1)$, $(2,2)$, and $(3,1)$. Most of the samples belong to orbit $(1,1,1,1)$, and only one samples fall under the orbit $(3, 1)$. If we select a maximum photon number per mode of $n=2$, the corresponding event $E_{4, 2}$ holds three out of the four orbits, and seven out of the eight samples. }\label{Fig:coarse-graining}
\end{figure}
\end{center}

Orbits present only a moderate coarse-graining
of the sample space: although there are many samples that fall into a given orbit, there remains
a large number of possible orbits. In fact, the number of possible orbits grows exponentially with the total number of photons~\cite{rademacher1938partition}. Therefore, finding the probability of an orbit can still be difficult experimentally, motivating a further
coarse graining. Such a method was suggested in
Ref.~\cite{bradler2019duality}. It consists of combining all samples with a total of $k$ photons such that the number of photons in any mode does not exceed a parameter $n$. We denote the set of all such samples as the \emph{event} $E_{k, n}$. For instance, the previous sample $S = (0, 2, 3, 0, 0, 0, 1, 2, 0)$ belongs to the event $E_{8, 3}$ but not to the event $E_{8, 2}$ since there is a mode where more than two photons were detected. In this sense, an event is simply a coarse-graining of orbits. This strategy for coarse-graining samples into orbits and events is illustrated in Fig.~\ref{Fig:coarse-graining}.
\vspace{-1cm}

Let $p_{k, n}(G)$ denote the probability of observing a sample belonging to the event $E_{k, n}$ when a graph $G$ has been encoded in a GBS device. As suggested in Refs.~\cite{bradler2018graph,schuld2019quantum, bradler2019duality}, a GBS feature map can be defined by mapping graphs to a corresponding feature vector consisting of event probabilities. More precisely, let $\bm{k} = \{k_{1},k_{2},\ldots,k_{D}\}$ be a vector specifying a range of total photon numbers and let $n$ be the maximum photon number per mode. The feature vector $\bm{f}_{\bm{k}, n}(G)$ assigned to the graph $G$ is given by
\begin{equation}
\bm{f}_{\bm{k}, n}(G) = \left(p_{k_{1}, n}(G), p_{k_{2}, n}(G), \ldots , p_{k_{D}, n}(G)\right).
\end{equation}
There is freedom in the choice of parameters $\bm{k}$ and $n$ defining the feature vector, as well as in the mean photon number of the GBS distribution. For any such choice, isomorphic graphs have the same feature vectors.
The complete GBS algorithm for constructing graph feature vectors is described below:

\vspace{0.3cm}

\begin{center}
\textit{Algorithm}
\end{center}
\begin{enumerate}
\item Encode the graph $G$ into the GBS device through its adjacency matrix $\bm{A}$ and generate $N$ samples from a distribution with mean photon number $\bar{n}$.
\item Given inputs $\bm{k}$ and $n$, for each of the $N$ samples, determine whether it belongs to any of the events $E_{k_i, n}$ for $i=1,2,\ldots,D$. Denote by $N_{i}$ the total number of samples that belong to the event $E_{k_i, n}$.
\item Compute the probability estimates $\tilde{p}_{k_{i}, n}(G)=\frac{N_i}{N}\approx p_{k_{i}, n}(G)$, where $p_{k_{i}, n}(G)$ is the probability of a sample belonging to the event $E_{k_{i}, n}$.
\item Assign the feature vector $\bm{f}_{\bm{k}, n}(G) = \left(\tilde{p}_{k_{1}, n}(G), \tilde{p}_{k_{2}, n}(G), \ldots , \tilde{p}_{k_{D}, n}(G)\right)$ to the graph $G$.
\end{enumerate}

GBS feature maps for graphs were tested in Ref.~\cite{schuld2019quantum} for a large collection of graph datasets~\cite{KKMMN2016}. The performance of GBS-based kernels in classification tasks was compared to three classical kernels: the Graphlet Sampling kernel~\cite{shervashidze2009efficient}, the random walk kernel~\cite{gartner2003graph}, and the subgraph matching kernel~\cite{kriege2012subgraph}. GBS kernels were computed exactly, and were found to outperform classical kernels in nine out of eleven datasets. 

\subsubsection{Using the GBS applications layer}

\begin{center}
\begin{figure}[t!]
\includegraphics[width= 0.24\columnwidth]{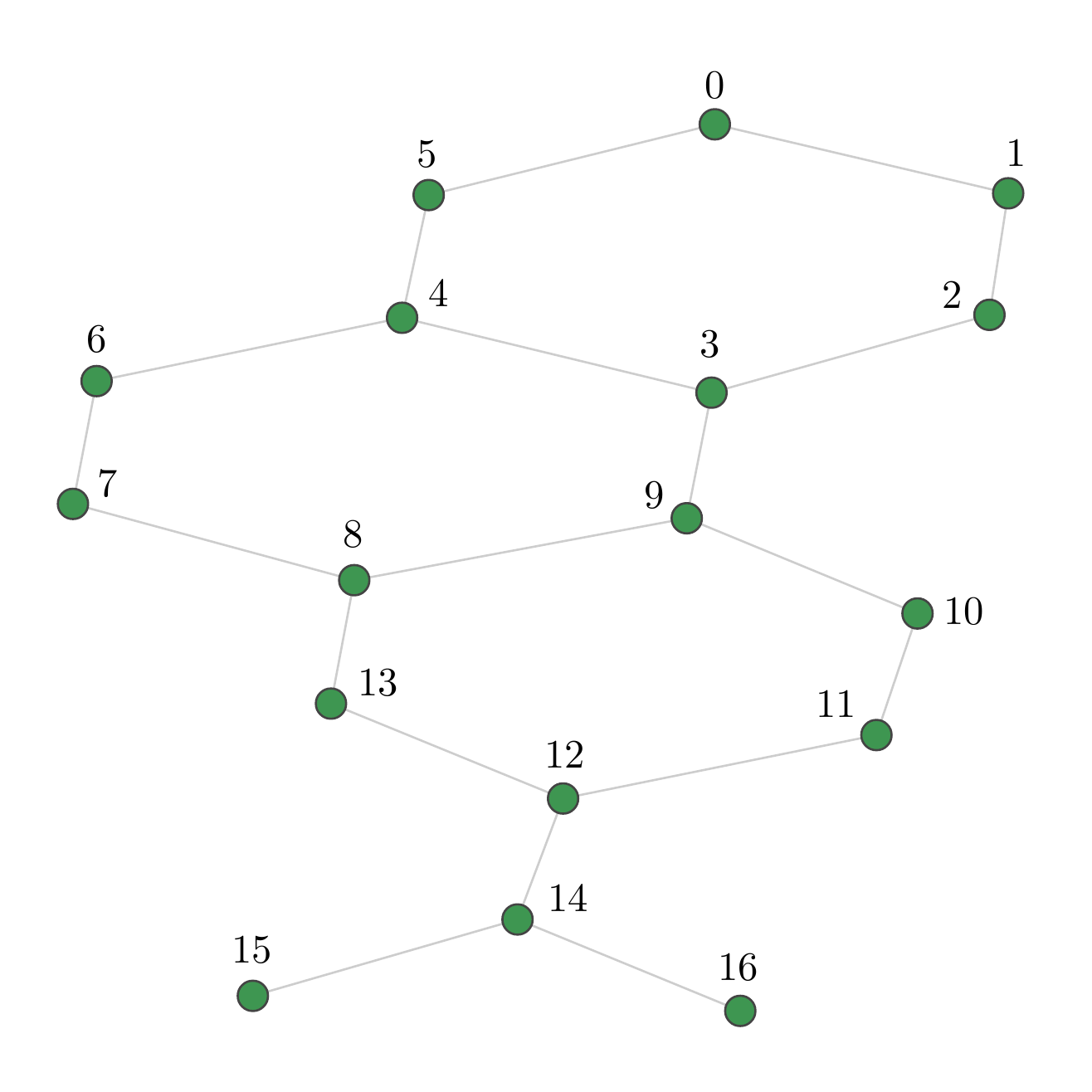}
\includegraphics[width= 0.24\columnwidth]{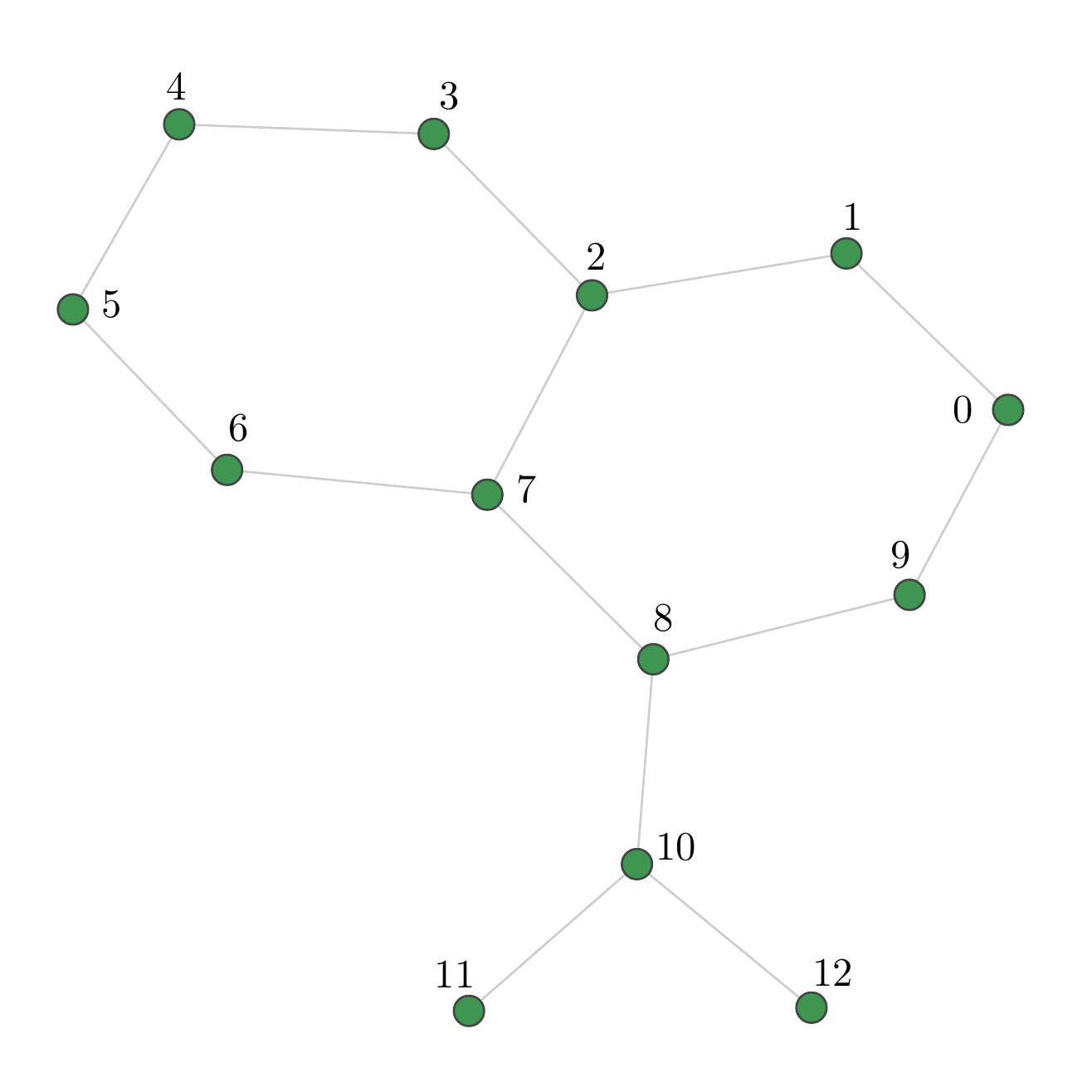}
\includegraphics[width= 0.24\columnwidth]{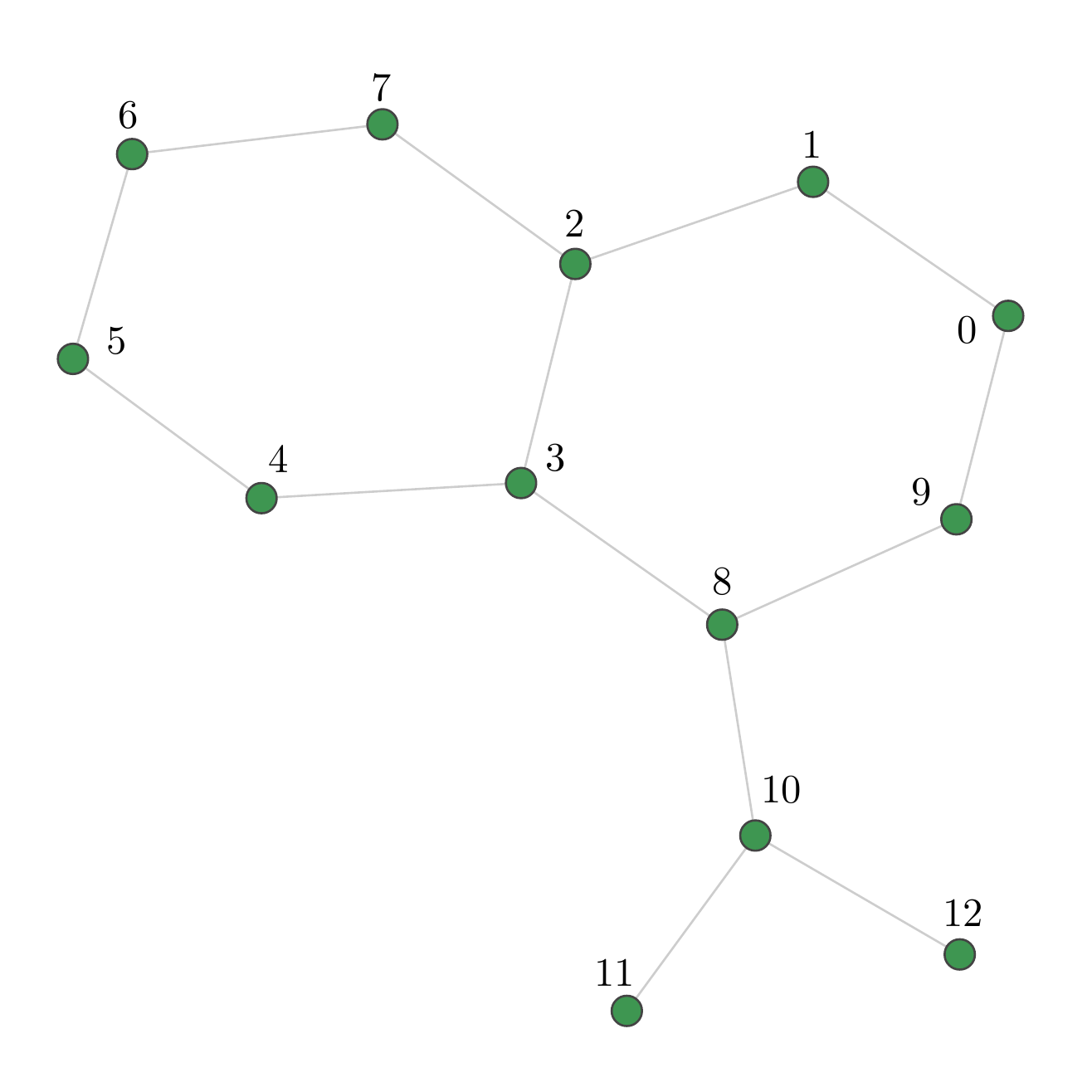}
\includegraphics[width= 0.24\columnwidth]{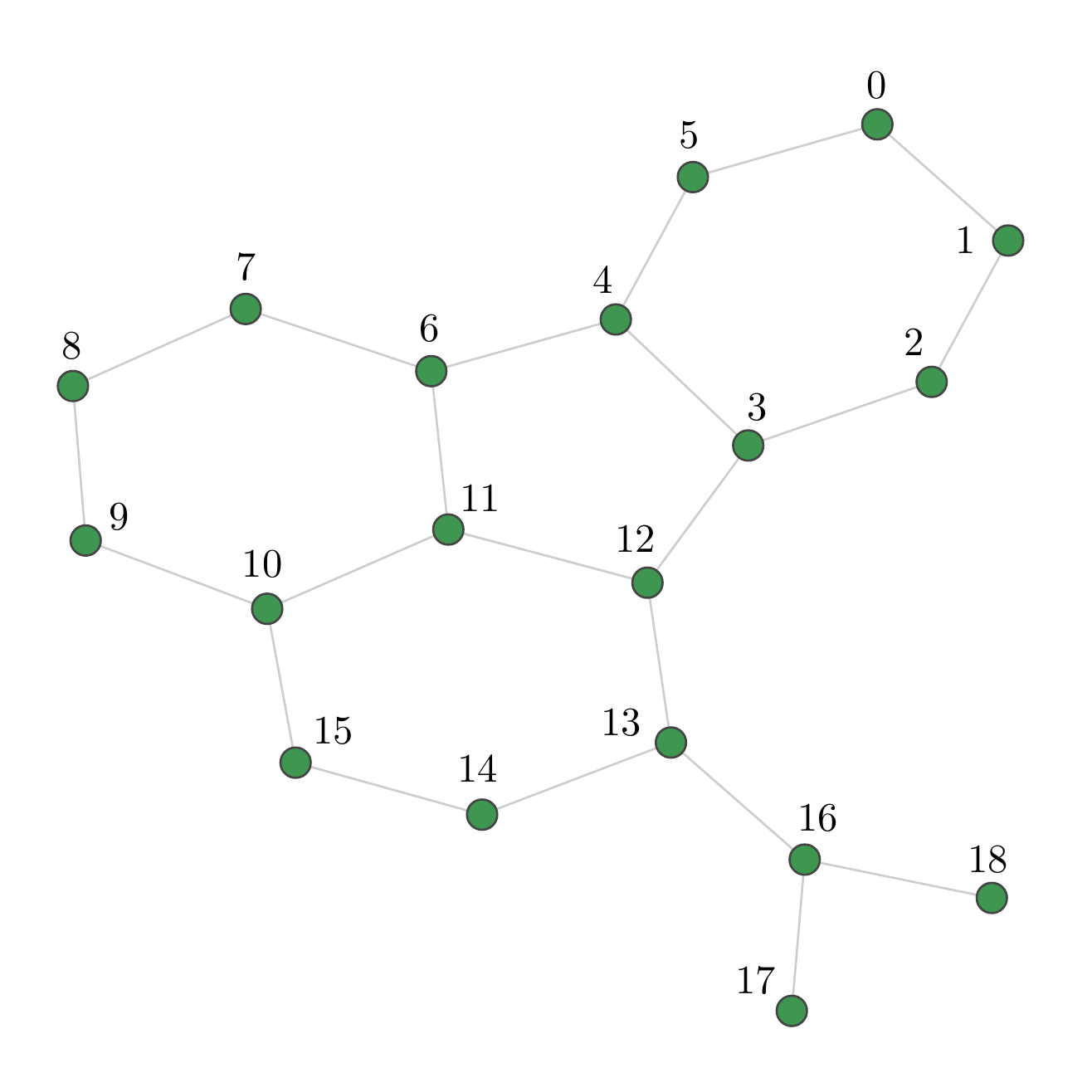}
\caption{The first four graphs of the MUTAG dataset~\cite{debnath1991structure,kriege2012subgraph}, including node labels.
From left to right, the first and last graph belong to the same mutagenic class, while the two graphs in the center both belong to a different class. Note that the middle graphs are isomorphic.}
\label{Fig:MUTAG_Graphs}
\end{figure}
\end{center}

As in previous sections, to demonstrate usage of the applications layer, we use pre-generated samples available in the \python{data} module. We focus on the MUTAG dataset, which consists of graphs representing the structure of a chemical compound~\cite{debnath1991structure,kriege2012subgraph}. Each graph is assigned a label based on its mutagenic effect. The \python{data} module provides four graphs from the MUTAG dataset: \python{Mutag0}, \python{Mutag1}, \python{Mutag2}, and \python{Mutag3}, each containing 20,000 samples from a distribution with mean photon number $\bar{n}=6$. The graphs are plotted in Fig.~\ref{Fig:MUTAG_Graphs}.

\vspace{-1cm}

The code below describes how feature vectors can be built for the above four MUTAG graphs. It employs the \python{similarity} module, which contains the function \python{feature_vector_sampling()} that can be used for constructing feature vectors from GBS samples. For plotting purposes, we focus on two-dimensional vectors defined by the events $E_{8, 2}$ and $E_{10, 2}$, specified by the vector $\bm{k}=(8, 10)$ and maximum number of photons per mode $n=2$. 

\begin{center}
\begin{figure}[t!]
\includegraphics[width= 0.45\columnwidth]{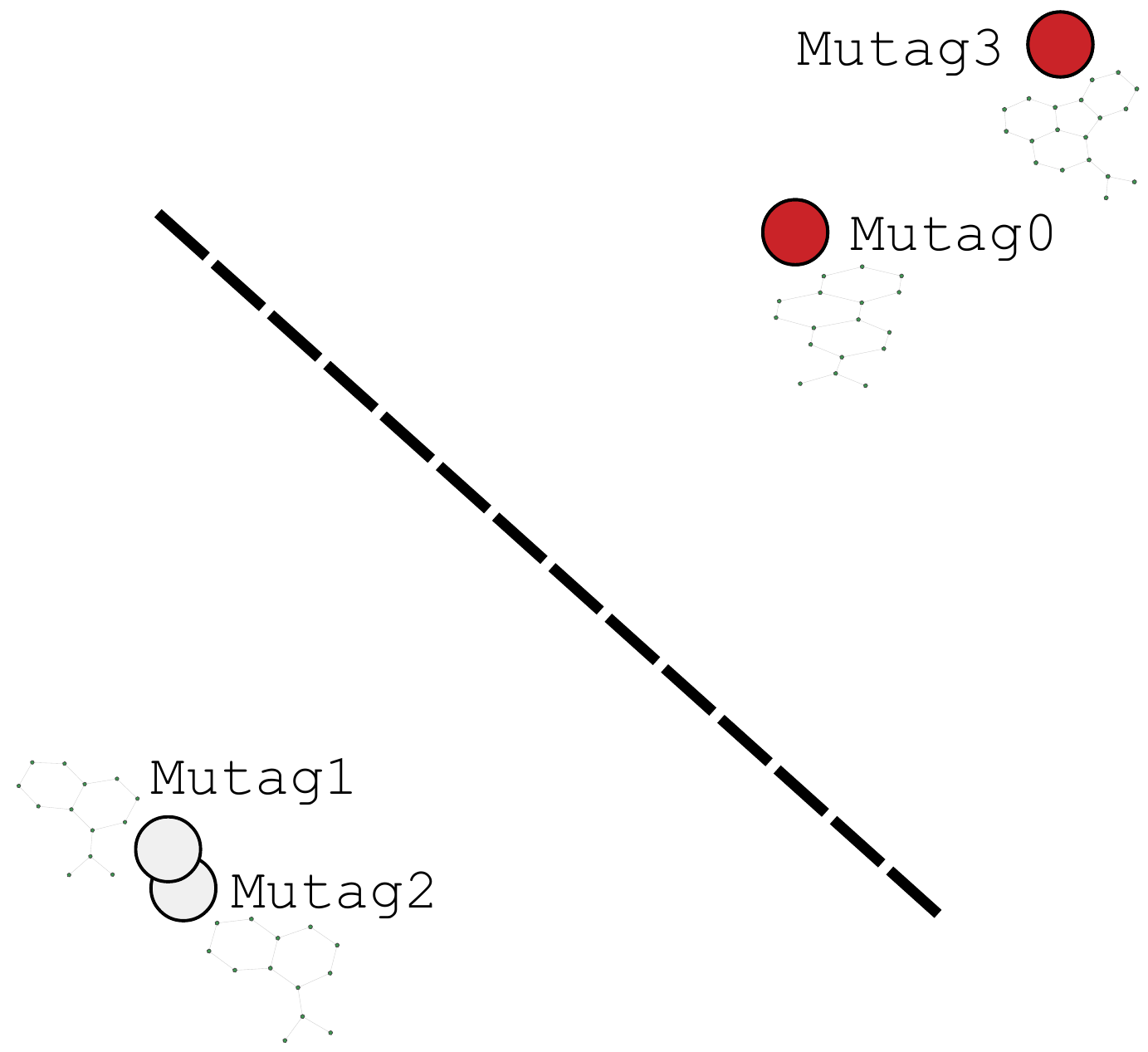}
\caption{A two-dimensional GBS feature space with the points each representing one of the
four MUTAG graphs. The red points correspond to one class and the light grey points to the other.
Here, the $x$-axis corresponds to the probability of event $E_{8, 2}$ and the $y$-axis
corresponds to the probability of event $E_{10, 2}$. A decision boundary given by training
scikit-learn’s Support Vector Machine \python{LinearSVC}~\cite{scikit-learn} on the data is given
as the black dashed line.}
\label{Fig:MUTAG}
\end{figure}
\end{center}

\begin{customcode}
>>> from strawberryfields.apps import data, similarity
>>> events = [8, 10]  # event photon numbers
>>> max_count = 2  # maximum number of photons per mode
>>> datasets = data.Mutag0(), data.Mutag1(), data.Mutag2(), data.Mutag3()
>>> f1, f2, f3, f4 = (similarity.feature_vector_sampling(data, events, max_count) for data in datasets)  # create feature vectors
\end{customcode}
The function \python{feature_vector_sampling()} directly computes the feature vector; it internally allocates samples into the corresponding events and keeps track of the number of samples detected for each event. In other words, feature vectors can be obtained from samples using a single line of code. The resulting two-dimensional feature vectors are shown in Fig.~\ref{Fig:MUTAG}. From this figure, it is noticeable that the points from the two isomorphic graphs are close to each other, the discrepancy arising from the statistical error in the estimation. The remaining graphs, which belong to another mutagenic class, are sufficiently further away to allow the construction of a linear decision boundary to classify both classes.

Finally, while accessing orbit or event probabilities can be feasible experimentally, numerically calculating these
probabilities is challenging, since typically a large number of sample probabilities must be evaluated. To test ideas before requesting access to hardware or using simulators, the applications layer also provides support for computing Monte Carlo estimates of orbit or event probabilities. A Monte
Carlo estimate of $p(E_{k, n})$ for the event $E_{k, n}$ is carried out by generating $N$ samples of $k$ photons $\{S_{1},S_{2},\ldots,S_{N}\}$ uniformly
at random from $E_{k, n}$ and evaluating
\begin{equation}
p(E_{k, n})\approx \frac{1}{N}\sum_{i=1}^N p(S_i) |E_{k, n}|,
\end{equation}
where $|E_{k, n}|$ is the number of samples in the event. However, being proportional to hafnians, the probabilities become exponentially difficult to compute for increasing $k$. The resulting feature vector can be calculated using the \python{feature_vector_mc()} function.

\subsection{Point processes}

Point processes are statistical models that generate random collections of points in a state space $\mathcal{M}$ according to a given probability distribution $\mathcal{P}$~\cite{daley2003introduction, daley2007introduction, baddeley2007spatial}.
We focus on point processes where $\mathcal{M}$ is the set of all non-negative integer-valued vectors $S=(s_1, s_2,\ldots, s_m)$ of dimension $m$. An important family of point processes are \emph{matrix} point processes. They are defined by a probability distribution $\mathcal{P}(S)$ given by
\beq
\mathcal{P}(S) \propto \Phi(\bm{K}_{S}),
\eeq
where $\bm{K}$ is a symmetric \emph{kernel matrix} and $\Phi$ is a matrix function such that $\Phi(\bm{K}_S)\geq 0$ for all matrices $\bm{K}_S$ selected by $S$. Specific classes of matrix point processes are determined by the choice of matrix function and, among each class, the properties of the resulting point process are uniquely determined by the choice of kernel matrix. One example class is the determinantal point process~\cite{borodin2009determinantal}, which is useful for generating diversity within the sampled points.

Point processes can be used to model stochastic data or as components in algorithms to solve specific tasks. For example, they can be used to provide seeds for training clustering algorithms~\cite{jahangiri2019point}. Point processes find applications in a variety of fields such as finance~\cite{bjork1997bond, bauwens2009modelling, embrechts2011multivariate, bacry2015hawkes}, biology~\cite{Li2016Point, Fromion2013Stochastic, Dodgson2013Spatial}, medicine~\cite{Jing2010Detection, Grell2015Three, Fok2012Functional}, physics~\cite{Picinbono2010Some, Klatt2014Characterization, Wohrer2019Ising, Schmidt2017Inertial}, chemistry~\cite{Halle2013Analysis, Persson2013Transient,Tzoumanekas2006Topological,Lazenby2018Quantitative,
Talattof2018Pulse}, and machine learning~\cite{kulesza2010structured, kulesza2011k, lin2012learning,kulesza2012determinantal,gillenwater2014expectation}. 

We now focus on the choice of kernel matrix. In spatial point processes, the state space is a set of points in $n$-dimensional space characterized by their coordinate vectors $\bm{r}_{1}, \bm{r}_{2}, \ldots, \bm{r}_{m} \in \mathbb{R}^{n}$. It is common to encode these points into a \emph{radial basis function} (RBF) kernel, defined as
\beq \label{Eq:RBFKernel}
\bm{K}_{ij}=e^{-\|\bm{r}_{i}-\bm{r}_{j}\|^2/2\sigma^2},
\eeq
where $\sigma$ is a parameter of the kernel and $\|\cdot\|$ is the Euclidean norm. The RBF kernel is positive semidefinite and represents a measure of similarity between points: those that are close to each other have large entries in the kernel matrix; points that are far away have correspondingly small entries. The role of $\sigma$ is to set the scale for determining proximity.

The RBF kernel can be modified in order to introduce additional control over the location of sampled points~\cite{jahangiri2019point}. Let 
$\vec{\lambda}=(\lambda_1, \lambda_2, \ldots, \lambda_m)$ be a vector assigning a weight $\lambda_i$ to the $i$-th point. A rescaled RBF kernel can then be defined as
\beq
K_{ij}=\lambda_i\lambda_je^{-\|\bm{r}_{i}-\bm{r}_{j}\|^2/2\sigma^2}.
\eeq
Finally, an alternative example of a positive semidefinite kernel is a correlation matrix, measuring pairwise correlations between random variables.

\subsubsection{GBS algorithm for point processes}

GBS devices can realize a physical implementation of a number of matrix point processes~\cite{jahangiri2019point}. For a
symmetric kernel
matrix $\bm{K}$ encoded according to the embedding discussed in
Sec.~\ref{Sec:Programming}, the device samples from a \emph{hafnian} point process
\beq\label{Eq:GBS-pp}
\Pr(S) \propto c^k \frac{|\text{Haf}(\bm{K}_S)|^{2}}{s_1!s_2!\ldots s_m!},
\eeq
where as in Eq.~\eqref{Eq: GBS_symm}, $c$ is the rescaling parameter that controls the mean photon number of the distribution and $k=\sum_i s_i$. Alternatively, any positive semidefinite $\bm{K}$ such as the RBF kernel can be encoded
using a variant of GBS
with thermal states as input~\cite{jahangiri2019point}, resulting in
a matrix point process known as a \emph{permanental} point process defined by the distribution
\beq
\Pr(S)\propto c^k\frac{\text{Per}(\bm{K}_S)}{s_1!s_2!\cdots s_m!}.
\eeq 

It is possible to use \emph{quantum-inspired} classical algorithms to efficiently sample from the
permanental point process with positive semidefinite kernel. The main insight is to employ the $P$ representation, where the state $\rho$ of a qumode is written as
\beq
\rho = \int d^2\alpha P(\alpha)\ket{\alpha}\bra{\alpha},
\eeq
where $\ket{\alpha}\bra{\alpha}$ denotes a coherent state and $P(\alpha)$ is a quasi-probability distribution. Thermal states have a positive $P$ function given by 
\beq\label{Eq:P-dbn}
P_{\text{th}}(\alpha, \mu)=\frac{1}{\pi \mu}\exp\left(-\frac{|\alpha|^2}{\mu}\right),
\eeq
where $\mu$ is the mean photon number of the state. A positive $P$ function implies that the state can be interpreted as a distribution over coherent states. Importantly, independent coherent states are mapped to independent coherent states by a linear interferometer. The photon-number distribution of a coherent state is Poissonian, which can be sampled efficiently. These features give rise to a polynomial-time quantum-inspired algorithm for permanental point processes, described below:
\begin{center}
\textit{Algorithm}
\end{center}
\begin{enumerate}
\item For an input positive semidefinite kernel matrix $\bm{K}$, compute its Takagi-Autonne decomposition $\bm{K}= \bm{U}\,\text{diag}(\mu_1,\mu_2,\ldots, \mu_m)\, \bm{U}^T$.
\item For a target mean photon number $\bar{n}$, solve for the parameter $c>0$ such that $\bar{n}=\sum_{i=1}^m\frac{c \mu_i}{1-c\mu_i}$.
\item For each mode $i=1,2,\ldots, m$, sample the parameter $\alpha_i \in \mathbb{C}$ from the distribution $P_{\text{th}}(\alpha_i, c\mu_i)$ of Eq.~\eqref{Eq:P-dbn}.
\item For each mode, compute the parameter $\beta_i=\sum_{j=1}^m U_{ji}\alpha_i$ and sample the integer $s_i\sim \text{Pois}(|\beta_{i}|^2)$ from a Poisson distribution with mean $|\beta_i|^{2}$.
\item Return the sample $S=(s_1, s_2, \ldots, s_m)$.
\end{enumerate}

This quantum-inspired algorithm for permanental point processes has an asymptotic runtime of $O(m^3)$, which is dominated by the requirement to compute the Takagi-Autonne decomposition of the input kernel matrix. Once this is done, generating a sample takes only $O(m^2)$ time. For all applications where the GBS input matrix is positive semidefinite, it is preferable to employ this fast quantum-inspired algorithm, which can scale to large input sizes.

The defining property of hafnian and permanental point processes is that, just like bosons bunching together, they sample \emph{clustered} collections of points. In spatial point
processes, these clusters are points that are close to each other, while point processes employing correlation matrices as kernels result in selecting highly-correlated variables.
This clustering property can potentially be used to model phenomena that generate clustered data, or it can be employed as a subroutine in algorithms aimed at identifying clusters, similarly to how GBS can accelerate algorithms for dense subgraph identification.

\subsubsection{Using the GBS applications layer}

The \python{points} module of the applications layer includes a function for computing the RBF kernel. The inputs are an $m\times n$ coordinate matrix $\bm{R}$---whose rows are the coordinate vectors of the points --- and the scaling parameter $\sigma$. The following example code shows how to construct a kernel for a $20\times 10$ rectangular array of points:
\begin{customcode}
>>> from strawberryfields.apps import points
>>> import numpy as np
>>> # define coordinate matrix
>>> R = np.array([(i, j) for i in range(20) for j in range(10)])
>>> sigma = 1.0  # kernel parameter
>>> K = points.rbf_kernel(R, sigma)  # creates kernel matrix
\end{customcode}

\begin{center}
\begin{figure}[t!]
\includegraphics[width= 0.8\columnwidth]{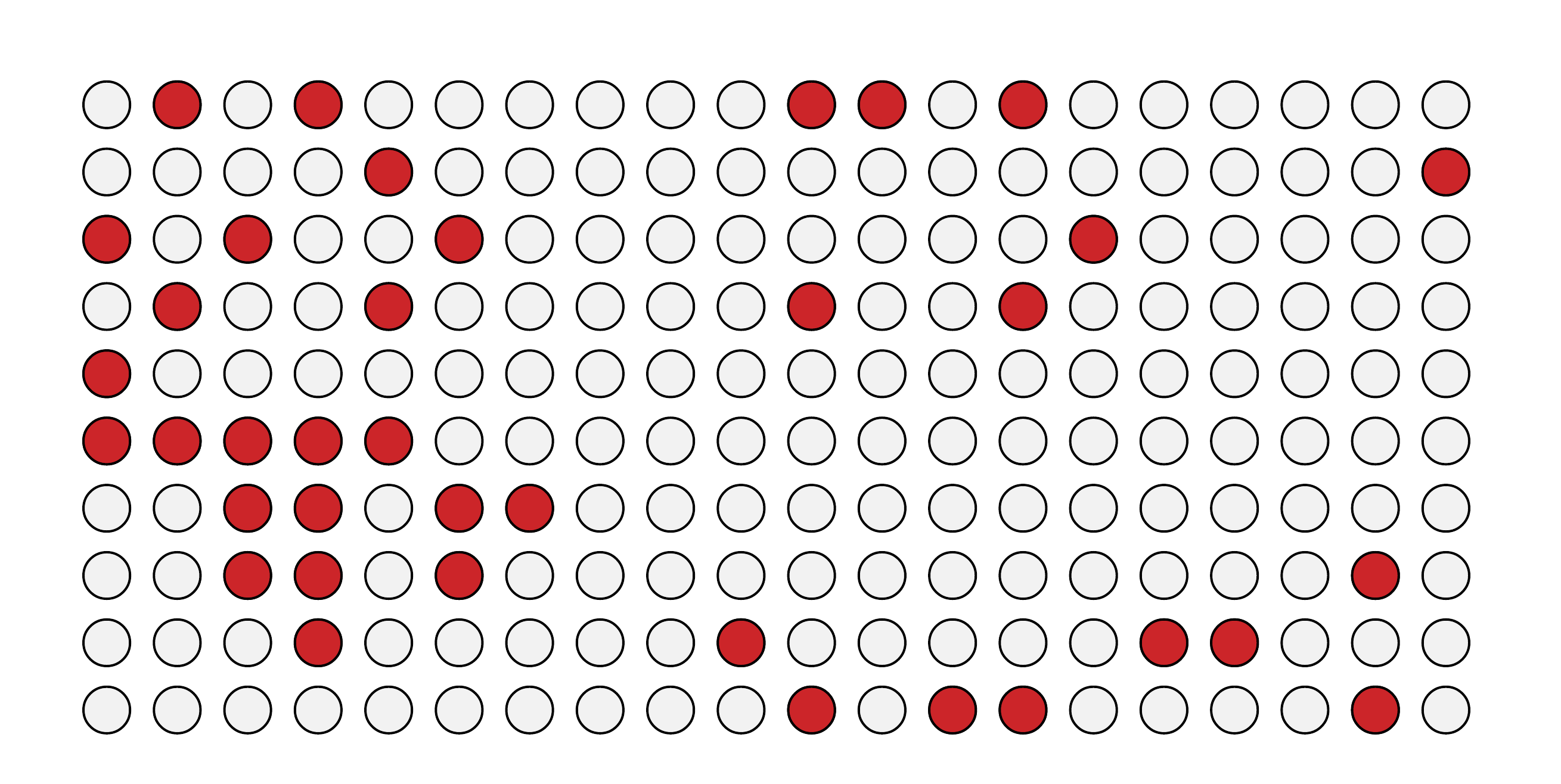}
\caption{Visualization of a sample from a permanental point process on a state space consisting of 200 points arranged in a $20\times 10$ rectangular grid. Sampled points are highlighted in red. Permanental and hafnian point processes are characterized by a clustering of the sampled points, as can be seen in the left of the figure.}\label{Fig:points}
\end{figure}
\end{center}

\vspace{-1.8cm}
A quantum-inspired algorithm can be used for the permanental point
process with the RBF kernel. Sampling for this case is performed in the applications layer with the
\python{sample()} function.
Continuing from the example code above, we can generate 1,000 samples with a mean photon number of 50:
\begin{customcode}
>>> n_mean = 50  # mean number of photons
>>> n_samples = 1000  # number of samples
>>> samples = points.sample(K, n_mean, n_samples)  # generates samples
\end{customcode}

The \python{plot} module includes built-in functionalities for visualizing point patterns. They can be accessed as follows, where we plot the first sample obtained:
\begin{customcode}
>>> from strawberryfields.apps import plot
>>> plot.points(R, samples[0], plot_size=(1000,500))  # See Fig. \ref{Fig:points}
\end{customcode}
Point processes are a particularly useful framework for visualizing GBS: the properties of the distribution can be directly showcased in terms of visible point patterns. In this particular sample, there is a cluster of points on the left of Fig.~\ref{Fig:points}. 

State spaces where all points are separated by an equal distance to their nearest neighbours are referred to as \emph{homogeneous state spaces}. More generally, points in state space may be distributed unevenly, in which case they are referred to as \emph{inhomogeneous state spaces}. In such scenarios, hafnian and permanental point processes encoded with an RBF kernel are more likely to sample points that are close together, i.e., points that already form a cluster.

\begin{center}
\begin{figure}[t!]
\includegraphics[width= 0.7\columnwidth]{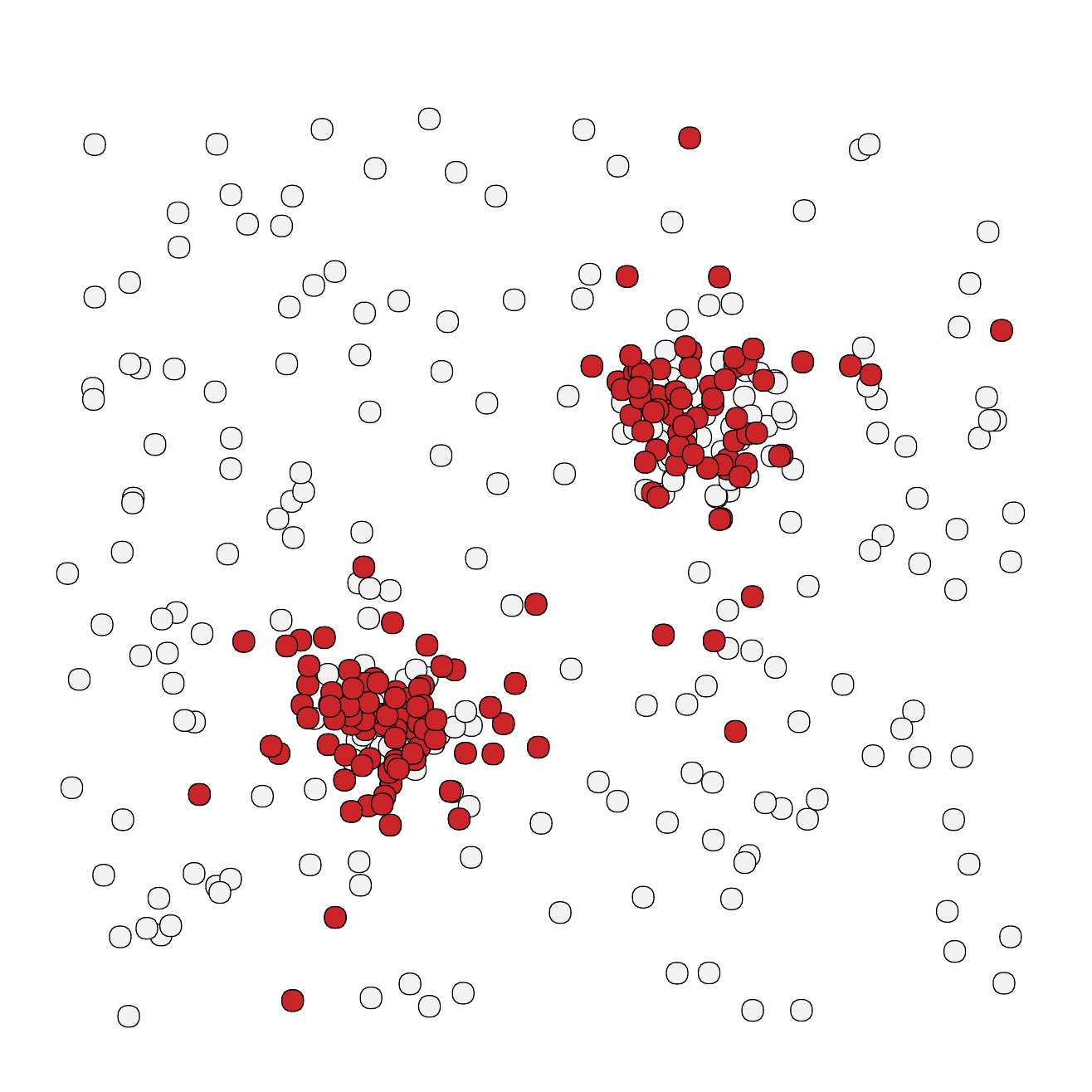}
\caption{A sample from a permanental point process on a state space consisting of two standard normal-distributed clusters of $100$ points and a uniform background of $200$ points. Sampled points are highlighted in red. They are more likely to occur towards the centers of the clusters.}\label{Fig:inh_points}
\end{figure}
\end{center}

In the example code below, we form a state space consisting of two clusters of $100$ points from the standard normal distribution, along with a background of $200$ uniformly distributed points. Samples are generated from the corresponding permanental point process. As shown in Fig.~\ref{Fig:inh_points}, the majority of selected points are close to the cluster centers, with few points selected from the background. This can be contrasted with a point process that selects points uniformly at random, where half of selected points would be expected to originate from the background.

\begin{customcode}
>>> cluster1 = np.random.normal(2, 0.3, (100, 2))
>>> cluster2 = np.random.normal(4, 0.3, (100, 2))
>>> background = np.random.rand(200, 2) * 6.0
>>> R = np.concatenate((cluster1, cluster2, background))
>>> sigma = 1.0  # kernel parameter
>>> K = points.rbf_kernel(R, sigma)  # creates kernel matrix
>>> n_mean = 50  # mean number of photons
>>> n_samples = 10  # number of samples
>>> samples = points.sample(K, n_mean, n_samples)  # generates samples
>>> plot.points(R, samples[0], point_size=10)  # See Fig. \ref{Fig:inh_points}
\end{customcode} 

\vspace{-1.5cm}

\subsection{Vibronic spectra}
Molecules absorb light at frequencies that depend on the allowed energy transitions between different states. These transitions can be determined both by electronic and vibrational degrees of freedom, in which case the absorption lines are referred to as the \emph{vibronic spectrum} of the molecule~\cite{sharp1964franck, doktorov1977dynamical, herzberg1988spectra, ruhoff2000algorithms, jankowiak2007vibronic, santoro2007effective}. The absorption
spectra of molecules is relevant for example in determining their usage in photovoltaics~\cite{hachmann2011harvard} or as dyes in industrial processes~\cite{gross2000improving}.

According to the Franck-Condon approximation~\cite{sharp1964franck}, the probability of an electronically-allowed vibronic transition is given by the \emph{Franck-Condon factor}. Starting from the ground state of a molecule at zero temperature---which corresponds to the vacuum of ground state vibrational excitations $\ket{\mathbf{0}}$---the Franck-Condon factors can be written as:

\beq
    F(\bm{m}) = \left | \bra{\bm{m}}\hat{U}_{\text{Dok}}\ket{\mathbf{0}} \right | ^ 2,
\eeq
where $\hat{U}_{\text{Dok}}$ is known as the \emph{Doktorov operator} and $\ket{\bm{m}}=\ket{m_1, m_2,\ldots, m_M}$ is the state denoting the number $m_i$ of phonons (a quantum of vibrational mechanical energy) in vibrational mode $i$ of the excited electronic state.

The vibronic spectrum is given by a \emph{Franck-Condon profile} (FCP), which determines the probability of generating a transition at a given vibrational frequency $\omega_{\text{vib}}$. At zero temperature, it is defined as
\beq
\text{FCP}(\omega_{\text{vib}})=\sum_{\bm{m}}F(\bm{m})\,\delta\left(\omega_{\text{vib}}-\sum_{k=1}^M m_k\omega'_k\right),
\eeq
where $\omega'_k$ is the frequency of the $k$-th vibrational mode of the excited electronic state and $\delta(\cdot)$ is the Dirac delta function.

At zero temperature, all vibronic transitions start from the ground vibrational state. However, at finite temperatures, the low-energy excited vibrational modes get populated and the vibronic transitions can start from excited vibrational states. To account for such transitions, the FCP at finite
temperature $T$ must include the initial distribution $P_T
(\bm{n})$ of the phonon modes in the electronic ground state, and also the energies of the initial vibrational modes. The FCP for finite-temperature vibronic transitions is then given by
\beq\label{Eq:FCP_finite}
\text{FCP}_{T}(\omega_{\text{vib}}) = \sum_{\bm{n}, \bm{m}}P_T
    (\bm{n})\left | \left
    \langle \bm{m} \left | \hat{U}_{Dok} \right | \bm{n} \right \rangle \right |^2  \delta \left(\omega_{\text{vib}} - \sum_{k=1}^{M}\omega'_km_k + \sum_{k=1}^{M}\omega_kn_k    \right),
\eeq
with $\omega_k$ the frequency of the $k$-th vibrational mode of the initial electronic state.

Several methods have been introduced to compute FCPs~\cite{ruhoff2000algorithms, jankowiak2007vibronic, santoro2007effective, quesada2019franck}, but it remains a significant challenge to accurately calculate them for large molecules~\cite{ezSpectrum}. The GBS algorithm we describe below is a method for leveraging the quantum properties of bosonic systems to efficiently reconstruct vibronic spectra.   

\subsubsection{Vibronic spectra with GBS}
\begin{center}
\begin{figure}[t!]
\includegraphics[width= 0.55\columnwidth]{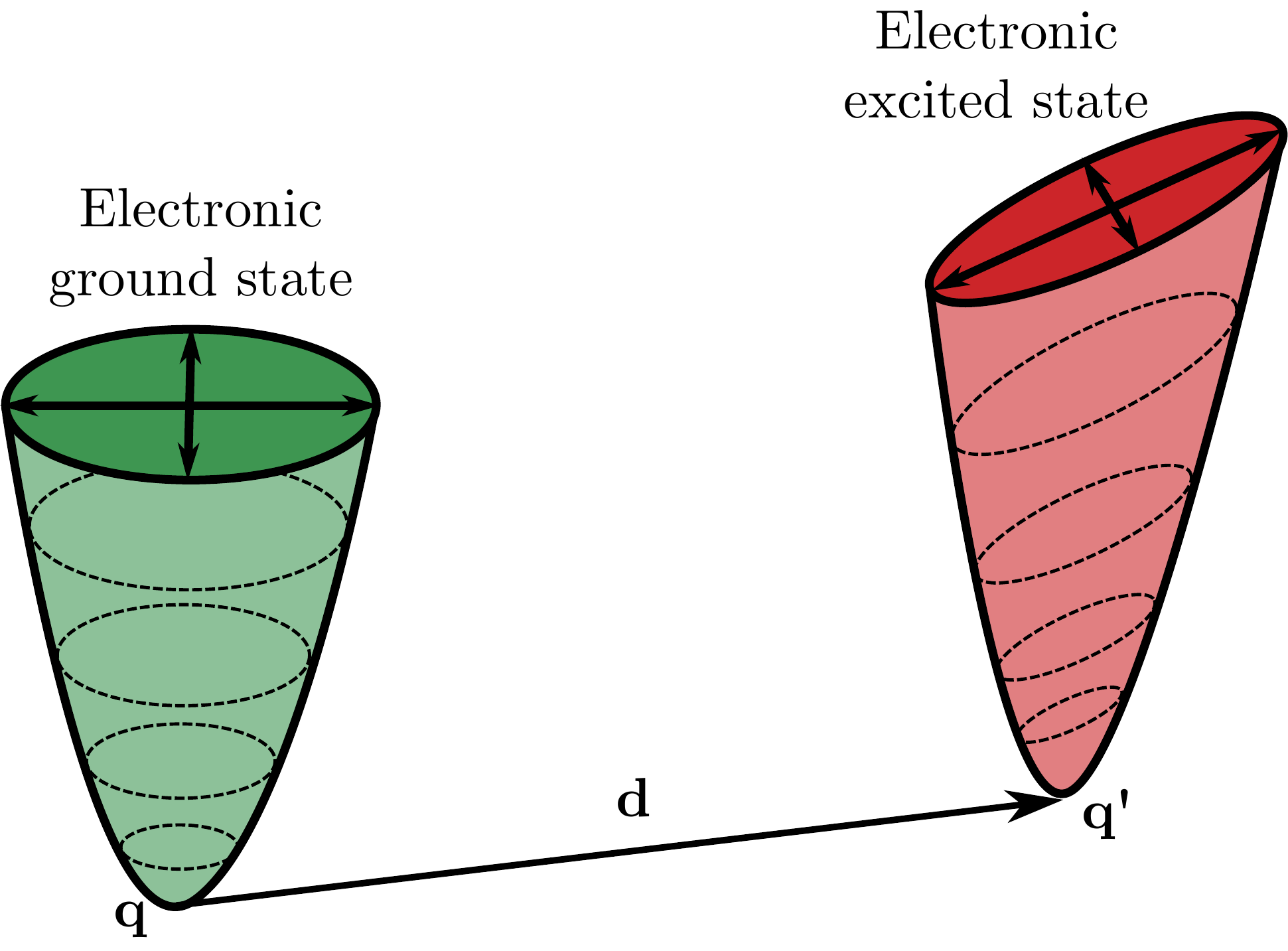}
\caption{Schematic illustration of the transformation connecting vibrational modes from the electronic ground and excited states. In the harmonic approximation, the energy surface of two vibrational modes can be represented by a paraboloid. The energy surface for the electronic ground state is shown in green on the left. The energy surface for the electronic excited state is shown in red on the right. The transformation linking the vibrational modes generally includes a rotation, squeezing, and displacement of the paraboloids. In the GBS algorithm for vibronic spectra, these operations are mapped to corresponding Gaussian transformations.   }\label{Fig:vibronic}
\end{figure}
\end{center}

It was shown in Ref.~\cite{huh2015boson} that a GBS device can be used to compute Franck-Condon profiles. Since then, there has been significant interest in generalizing the original techniques~\cite{huh2017vibronic, clements2018approximating, quesada2019franck} and in performing related experiments~\cite{sparrow2018simulating}. The main insight underlying the connection between GBS and vibronic spectra is the mathematical equivalence between photons in an optical mode and phonons in a vibrational mode. This is particularly showcased by the fact that the Doktorov operator can be expressed as a Gaussian unitary, which can be decomposed in terms of multi-mode displacement $\hat{D}(\bm{\alpha})$, squeezing
$\hat{S}(\bm{\Sigma})$, and linear interferometer $\hat{R}(\bm{C}_L)$, $\hat{R}(\bm{C}_R)$ operators as~\cite{quesada2019franck}:

\beq\label{Eq: Dok_Gaussian}
    \hat{U}_{Dok} = \hat{D}(\bm{\alpha})\hat{R}(\bm{C}_L) \hat{S}(\bm{\Sigma}) \hat{R}(\bm{C}_R),
\eeq
where $\bm{C}_L$ and $\bm{C}_R$ are unitary matrices, $\bm{\Sigma}$ is a diagonal matrix, and $\bm{\alpha}$ is a vector. We describe these parameters in more detail below. The Doktorov operator describes a transformation between the vibrational modes of the ground and excited electronic states, which is illustrated in Fig.~\ref{Fig:vibronic}.

The full Franck-Condon profile can be statistically reconstructed by preparing the input modes in a thermal state, applying the Doktorov transformation, and measuring the output modes in the Fock basis to generate a histogram of the sampled energies. The peaks of this histogram correspond to the absorption lines of the vibronic spectrum of the molecule. Thermal states can be prepared as the reduced states of two-mode squeezed states with squeezing parameter $r_i$ given by
\beq\label{Eq:2mode_squeeze}
r_i(\omega_i, T) = \arctanh\left(e^{-\hbar \omega_i/2k_B T}\right).
\eeq

The parameters $\bm{C}_L$, $\bm{C}_R$, $\bm{\Sigma}$, and $\bm{\alpha}$ that determine the Doktorov operator from Eq.\eqref{Eq: Dok_Gaussian} are obtained as follows. First, diagonal matrices $\bm{\Omega}$ and $\bm{\Omega'}$ are constructed respectively from the ground and excited state frequencies:

\begin{align}
\bm{\Omega} &= \text{diag} (\sqrt{\omega_1},...,\sqrt{\omega_k}),\\
\bm{\Omega'} &= \text{diag} (\sqrt{\omega_1'},...,\sqrt{\omega_k'}).
\end{align}
We also define the \emph{Duschinsky matrix} $\bm{U}$, which is obtained from the normal mode coordinates of the ground and excited states, $\bm{q}$ and $\bm{q}'$ respectively, as
\beq
    \bm{q}' = \bm{U}\bm{q} + \bm{d}.
\eeq
The displacement vector $\bm{d}$ is related to the structural changes of the molecule upon vibronic excitation. Since these vectors are real, the unitary $\bm{U}$ has real entries, i.e., it is an orthogonal matrix. Furthermore, we define the matrix $\bm{J}$ as
\beq\label{Eq: J_matrix}
\bm{J}=\bm{\Omega'}\bm{U}\bm{\Omega}^{-1}.
\eeq

The matrices $\bm{C}_L$, $\bm{C}_R$, and $\bm{\Sigma}$ are obtained from the singular value decomposition of $\bm{J}$:
\beq\label{Eq:J_SVD}
\bm{J} = \bm{C}_L\bm{\Sigma}\bm{C}_R.
\eeq
Finally, the displacement vector $\bm{\alpha}$ is given by 
\beq\label{Eq:alpha_delta}
\bm{\alpha}=\bm{\delta}/\sqrt{2},
\eeq
with $\bm{\delta}=\hbar^{-1/2}\bm{\Omega'}\bm{d}$. All the elements are now in place to specify the GBS algorithm for vibronic spectra. Given inputs, $\bm{U}$, $\bm{d}$, $\bm{\Omega}$, $\bm{\Omega'}$, and temperature $T$, we proceed as follows:
\begin{center}
\textit{Algorithm}
\end{center}
\begin{enumerate}

\item Compute the parameters $\bm{C}_L$, $\bm{C}_R$, $\bm{\Sigma}$, and $\bm{\alpha}$ as in Eqs.~\eqref{Eq: J_matrix}, \eqref{Eq:J_SVD} and \eqref{Eq:alpha_delta}.
\item For each $i=1,2,\ldots, M$, where $M$ is the number of vibrational modes, prepare a two-mode squeezed state between modes $i$ and $i+M$ of a $2M$-mode GBS device. The squeezing parameter of each two-mode squeezed state is given by $r_i(\omega_i, T)$ as in Eq.~\eqref{Eq:2mode_squeeze}. 
\item Apply the Doktorov transformation of Eq.~\eqref{Eq: Dok_Gaussian} to the first $M$ modes of the device. This is done by sequentially applying the transformations $\hat{R}(\bm{C}_R)$, $\hat{S}(\bm{\Sigma}) $, $\hat{R}(\bm{C}_L) $, and $\hat{D}(\bm{\alpha})$.

\item Measure all output modes and denote the pattern of detected photons by $S=(\bm{m}; \bm{n})=(m_1, m_2, \ldots, m_M; n_1, n_2,\ldots, n_M)$. Compute the transition energy 
\beq\label{Eq:energy_sample}
E(S)=\sum_{k=1}^{M}\omega'_km_k - \sum_{k=1}^{M}\omega_kn_k,
\eeq
defined by this output.
\item Generate $N$ samples in this manner and create a histogram for all observed energies.
\end{enumerate}
In the specific case of zero temperature, all the two-mode squeezing parameters are zero, and the input state to the Doktorov transformation is the vacuum. Since an interferometer maps the vacuum to itself, it is then possible to forego the two-mode squeezing and the first interferometer, leading to a simpler circuit on only $M$ modes.

\subsubsection{Using the GBS applications layer}

The applications layer contains the \python{vibronic} module
for computing vibronic spectra. Given the input chemical parameters $\bm{U}$, $\bm{d}$, $\bm{\Omega}$, $\bm{\Omega'}$, and $T$, the first step is to calculate the GBS parameters. This can be done using the \python{gbs_params()} function. Following Ref.~\cite{huh2015boson}, here we focus on the vibronic spectrum of formic acid. The chemical parameters for formic acid, as well as pre-generated samples, are available in the \python{data} module:
\begin{customcode}
>>> from strawberryfields.apps import data
>>> formic = data.Formic()  # load data
>>> w = formic.w  # ground state frequencies
>>> wp = formic.wp  # excited state frequencies
>>> Ud = formic.Ud  # Duschinsky matrix
>>> delta = formic.delta  # displacement vector
>>> T = 0  # temperature
\end{customcode}

The GBS parameters are obtained by calling \python{gbs_params()}:

\begin{customcode}
>>> from strawberryfields.apps import vibronic
>>> t, U1, r, U2, alpha = vibronic.gbs_params(w, wp, Ud, delta, T)
\end{customcode}

The \python{sample} module contains the function \python{vibronic()} that is tailored for vibronic spectra sampling. It can be used to generate samples directly from the GBS parameters just computed:

\begin{customcode}
>>> from strawberryfields.apps import sample
>>> nr_samples = 2
>>> s = sample.vibronic(t, U1, r, U2, alpha, nr_samples)
\end{customcode}

\begin{center}
\begin{figure}[t!]
\includegraphics[width= 0.65\columnwidth]{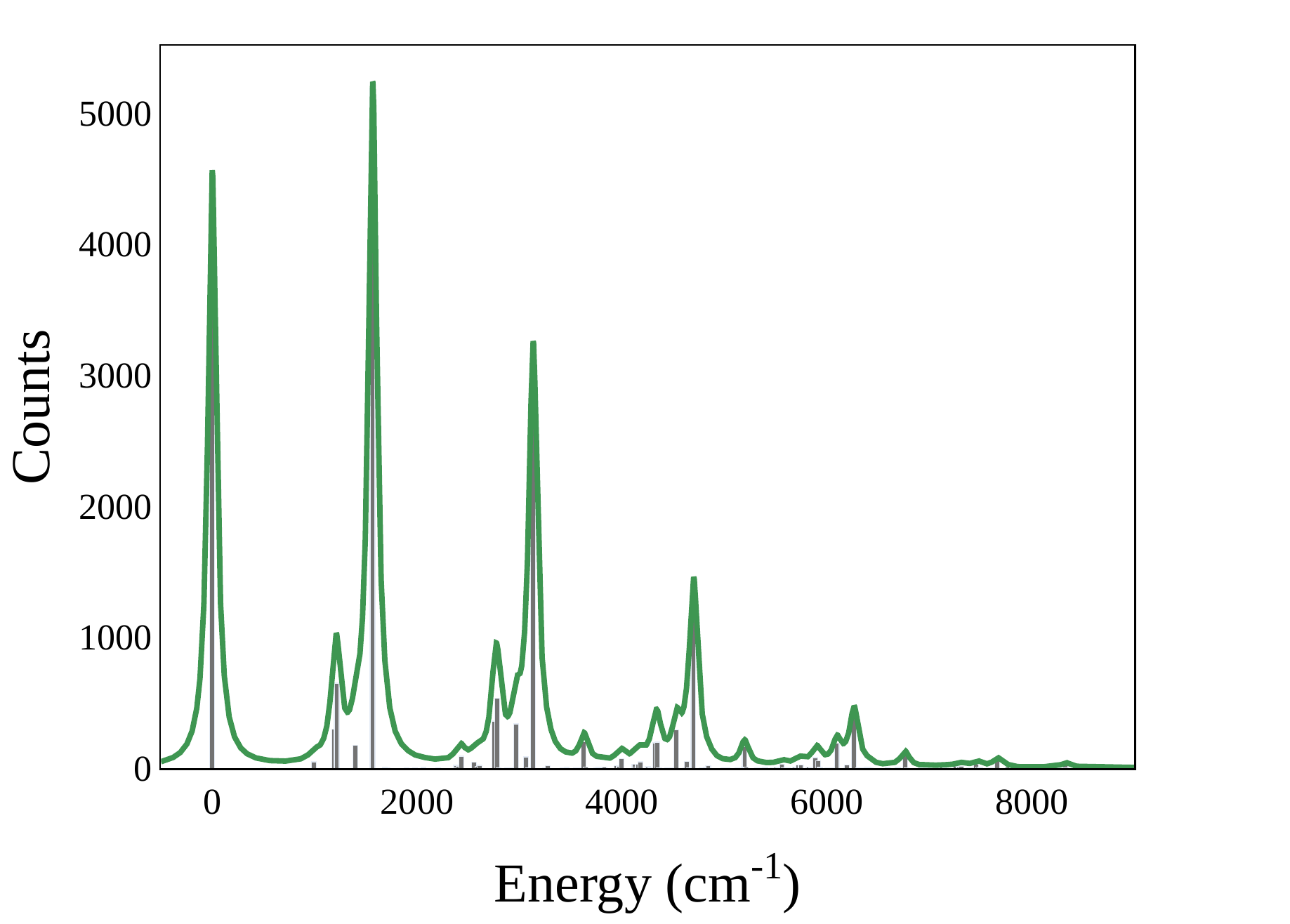}
\caption{Vibronic spectrum for formic acid reconstructed from a GBS algorithm. The vertical axis represents the number of samples observed. The horizontal axis is the relative wavenumber in cm$^{-1}$ of the corresponding transition, which, as is typically the case in spectroscopy, we interpret as an energy. Note that the wavenumber of the 0-0 transition is considered as zero relative wavenumber, and all the others have been scaled accordingly. Histogram bars are shown in grey, while a Lorentzian broadening of these bars is shown as the green curve above, which more closely represents experimental spectra. }\label{Fig:vib_spectrum}
\end{figure}
\end{center}

Besides molecular information, we can also load pre-generated samples from the \python{data} module and use the \python{energies()} function from the \python{vibronic} module to compute energies as in Eq.~\eqref{Eq:energy_sample}:
\begin{customcode}
>>> e = vibronic.energies(formic, w, wp)
\end{customcode}
Finally, the \python{plot} module contains a dedicated function, \python{spectrum()}, that can be used to construct the vibronic spectrum from the sampled energies:
\begin{customcode}
>>> from strawberryfields.apps import plot
>>> plot.spectrum(e, xmin=-500, xmax=9000) # See Fig. \ref{Fig:vib_spectrum}
\end{customcode}
The plot function builds a histogram of the energies and constructs a curve showing a Lorentzian broadening of the sampled energies. This mimics the physical broadening of the spectrum that is observed in practice.

\section{Outlook}
The goal of building quantum devices that are capable of performing previously inaccessible computations is that this computational power, however specific, can be harnessed to perform useful tasks. GBS has emerged as a platform to potentially unlock new computational capabilities in quantum photonics, and several recent works have described how the devices can be programmed to carry out algorithms in a wide range of applications. 

Significant work lies ahead on several fronts. First, not only is it crucial to discover new GBS algorithms, it is also important to connect known techniques to more specific, industrially-relevant problems of interest. Arguably the only such examples to date are the algorithm for vibronic spectra and the GBS algorithm for molecular docking reported in Ref.~\cite{banchi2019molecular}. In all cases, further work is also required to better understand the potential advantages of these methods compared to purely classical algorithms, particularly in determining scaling with respect to increasing problem size.  

Experimental implementations are in their infancy, but rapid progress is being made to develop these platforms. Inevitably, these will suffer from the presence of imperfections and decoherence, notably photon loss, which will affect the performance of the devices. It is therefore of utmost importance to fully grasp the impact of imperfections, to develop noise-resilient algorithms, and to conceive new experimental and theoretical methods for error mitigation. 

The Strawberry Fields applications layer is open-source software that was built to support and enable progress in all these fronts. It is not a final product, but a platform that is designed to grow and adapt as new discoveries are made, unlocking a virtuous cycle where quantum software aids quantum computing research, which in turn helps to build better software.  

\section*{Acknowledgements}
We thank Kamil Br\'adler, Ish Dhand, and Christian Weedbrook for helpful discussions.

\bibliographystyle{apsrev}
\bibliography{references}

\end{document}